\documentclass[fleqn,usenatbib]{mnras}

\usepackage{newtxtext,newtxmath}

\usepackage[T1]{fontenc}

\DeclareRobustCommand{\VAN}[3]{#2}
\let\VANthebibliography\thebibliography
\def\thebibliography{\DeclareRobustCommand{\VAN}[3]{##3}\VANthebibliography}

\usepackage{graphicx}	
\usepackage{amsmath}	
\UseRawInputEncoding

\title[Thermal evolution of lava planets]{Thermal evolution of lava planets}

\author[Herath et al.]{
Mahesh Herath$^{1,2}$\thanks{E-mail: mahesh.herath@mail.mcgill.ca}
Charles-\'Edouard Boukar\'e$^{3,4}$
Nicolas B. Cowan$^{1,2,5}$
\\
$^{1}$Department of Earth \& Planetary Science, McGill University, 3450 Rue University, Montreal, QC, Canada, H3A 0E8 \\
$^{2}$Trottier Space Institute, McGill University, 3550 Rue University, Montreal, QC, Canada, H3A 2A7 \\
$^{3}$Institut de Physique du Globe de Paris, 1 Rue Jussieu, Paris, France\\
$^{4}$Department of Physics \& Astronomy, York University, 140 Campus Walk Room 128, North York, Toronto, ON, Canada\\
$^{5}$Department of Physics, McGill University, 3600 Rue University, Montreal, QC, Canada\\
}

\date{Accepted XXX. Received YYY; in original form ZZZ}

\pubyear{2015}

\begin{document}
\label{firstpage}
\pagerange{\pageref{firstpage}--\pageref{lastpage}}
\maketitle

\begin{abstract}
Rocky planets are thought to form with a magma ocean that quickly solidifies. For lava planets, however, it is a permanent feature. The horizontal and vertical extent of this magma ocean depends on the interior thermal evolution of the planet, and possibly exogenous processes such as planet migration. We present a model for simulating the thermal history of tidally locked lava planets. We initiate the model with a completely molten mantle and evolve it for ten billion years. We adopt a fixed surface temperature of 3000 K for the irradiated day-side, but allow the night-side temperature to evolve along with the underlying layers. We simulate planets of radius 1.0$R_{\oplus}$ and 1.5$R_{\oplus}$ with different core mass fractions, although the latter does not significantly impact the thermal evolution. We confirm that the day-side magma ocean on these planets has a depth that depends on the planetary radius. The night-side, on the other hand, begins crystallizing within a few thousand years and completely solidifies within 800 million years in the absence of substantial tidal heating or day-night heat transport. We show that a magma ocean can be sustained on the night-side of a lava planet if at least 20 per cent of absorbed stellar power is transmitted from the day-side to the night-side via magma currents. Such day-night transport could be sustained if the magma has a viscosity of $10^{-3}$ Pa s, which is plausible at these temperatures. Alternatively, the night-side could remain molten if the mush layer is tidally heated at the rate of $8 \times 10^{-4}$ W/kg of mush, which is plausible for orbital eccentricities of $e > 7 \times 10^{-3}$. Night-side cooling is a runaway process, however: the magma becomes more viscous and the mush solidifies, reducing both day-night heat transport and tidal heating. Measurements of the night-sides of lava planets are therefore a sensitive probe of the thermal history of these planets. 
\end{abstract}



\section{Introduction}  \label{sec1}

Lava planets are terrestrial planets with ultra-short orbital periods. Examples include CoRoT-7b \citep{Leger2009}, Kepler-10b \citep{Batalha2011}, TOI-561b \citep{Patel2023} and K2-141b \citep{Barragan2018, Malavolta2018}. A lava planet's day-side is hot enough to melt and vapourize rock \citep{Schaefer2009}, while its night-side remains dark and cold if the planet is tidally locked into synchronous rotation. A lava planet therefore has two hemispheres experiencing different thermal evolutionary paths in parallel. 

Thermal phase curves of lava planets constrain their day-side and night-side temperatures. With such data we may be able to infer the history of lava planets by linking surface properties to an evolving interior structure and internal thermodynamics. To predict the characteristics of the magma ocean, we must create a thermal history model to simulate how lava planets evolve from their early days to the present. 

The magma oceans on lava planets are analogous to those that existed in the early Solar System \citep{Elkins2012, solo2000}. Where they differ is in their boundary conditions: the magma ocean on a lava planet is heated from above, and may be in a steady-state. Depending on where the geotherm of the mantle meets the solidus and liquidus for silicates, the mantle could also have a mush layer (melt fraction between 30 and 60 per cent). Models of magma oceans have been developed for planets in the Solar System \citep{Elkins2012} and applied to lava planets \citep{chao2021, Leger2011, Kite2016, Nguyen2020} predicts shallow day-side magma oceans. \cite{Boukare2022}, however, predicted deep magma oceans including extended mush zones. 

In this paper we develop a thermal model to simulate how a lava planet would evolve from a completely molten state to its present state. We use a pair of coupled 1D models for the two sides of a tidally locked planet to track how their interior structure and thermal profiles evolve. With this model we simulated the thermal evolution of lava planets of different sizes, core mass fractions (CMF), tidal heating and day-to-night heat transport. We describe our model in Section \ref{sec2}, present our results in Section \ref{sec3}, and discuss them in Section \ref{sec4}.

\section{Model}  \label{sec2}

\subsection{Single-column model}  \label{sec2.1}

\begin{figure*}\centering
\includegraphics[width=1.1\textwidth]{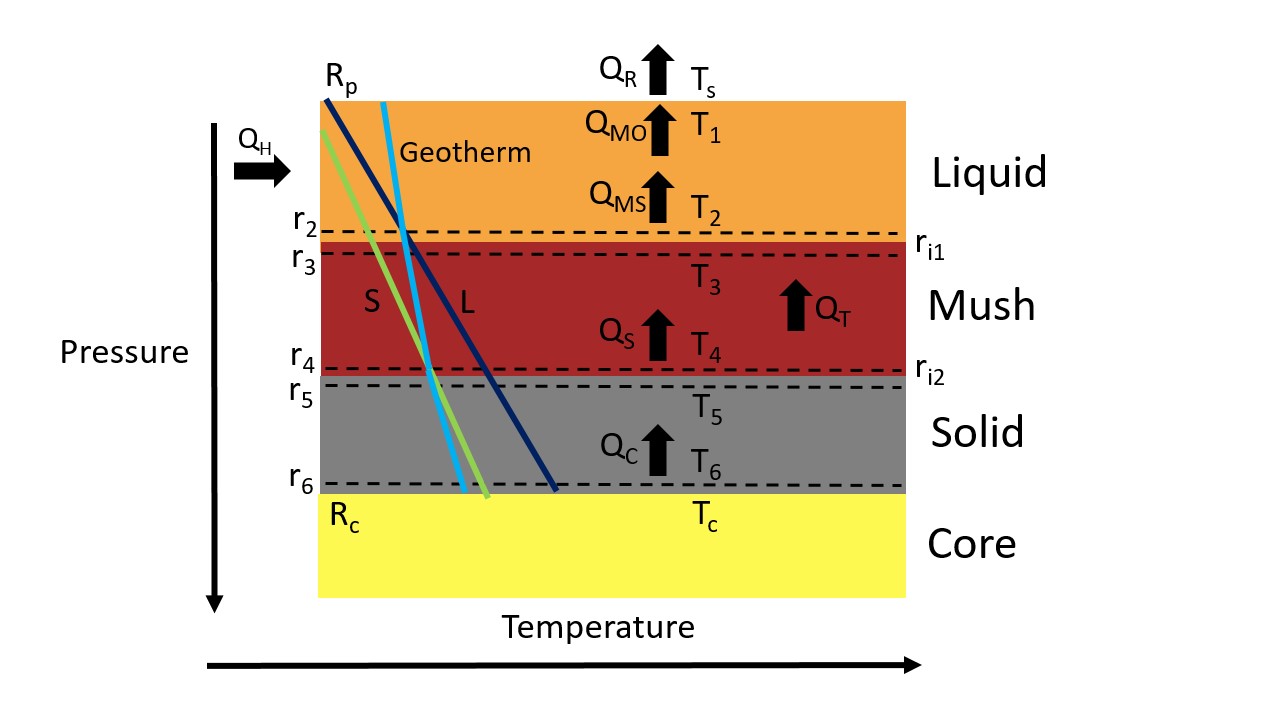}

\caption{Schematic of the 1D thermal evolution model used in this work. $Q_{R}$ is the radiated flux, $Q_{MO}$ is the heat flux from the magma ocean, $Q_{MS}$ the heat flux from the magma mush, $Q_{S}$ the flux from solid rock and $Q_{c}$ the flux from the core. $Q_{H}$ is the heat removed from the day-side and added to the night-side when we include horizontal heat transport. $Q_{T}$ is the tidal heating when tidal dissipation is included in the model. The light blue line is the geotherm of the mantle while the line "L" (dark blue line) is the mantle liquidus and "S" (green line) is the mantle solidus. The pressure increases in the downward direction while the temperature increases to the right.}
\label{fig1}
\end{figure*}

We considered a tidally locked planet with an Earth-like composition of peridotite (olivine and pyroxene). We additionally assumed that any existing atmosphere is optically thin \citep{Castan2011, Nguyen2020, Nguyen2022}. We approximated the planet's day-side surface temperature $T_{s}$ with the local equilibrium temperature at the sub-stellar point. Surface temperatures above 1800 K are necessary to sustain a molten surface \citep{Kite2016}. We used the sub-stellar temperature and radius of K2-141b to test out our models for Super-Earths, since K2-141b is the highest signal-to-noise ratio lava planet yet found \citep{Barragan2018, Malavolta2018}. It is also the subject of two approved James Webb Space Telescope observing campaigns \citep{Dang2021, Espinoza2021}. $T_{s}$ was set to 3000 K on the day-side, the sub-stellar temperature expected for K2-141 b \citep{Malavolta2018}. We neglect the physics of solid-liquid phase separation and focus only on the thermal aspect of magma ocean evolution. For the night-side the radiative equilibrium temperature is formally zero, so we allowed $T_{s}$ to vary in accordance with the flux escaping the surface. In the initial time step we started off with a value of $T_{s} = 3000$ on the night-side. In subsequent time steps, $T_{s}$ was found using the Stefan-Boltzmann law on flux radiated from the surface. 

As shown in Figure \ref{fig1}, we modelled the thermal evolution of this planet with a 1D parameterized model coupling silicate layers of different rheological properties with the core. Initially the mantle was assumed to be fully molten, meaning that at the start of the simulation we have the core and a layer of magma. The cooling of the magma ocean (MO) would lead to the formation of magma mush (MS) followed by the formation of solid rock (S) in the mantle. The formation of mush and solid rock was determined by the intersection of the geotherm with the liquidus and solidus curves for bulk silicate material. While the radial position of the core-mantle-boundary (CMB) remains unchanged, the number of layers in the mantle as well as their respective thickness change with time. We applied a viscosity ($\eta$) of $\eta$ = 0.1 Pa s (a melt fraction greater than 50 per cent) for the magma \citep{Xie2021, Losq2023}. We assumed the the mush is composed of  50 \% of solid. The viscosity of the partially molten mushy layer was set to 10$^{14}$ Pa s (a melt fraction between 10 and 40 per cent)  \citep{Xie2021, Losq2023}. For solid rock we used $\eta$ = 10$^{18}$ Pa s \citep{Ty2015}. As shown in Figure \ref{fig1} the parameters $T_{1}$ to $T_{6}$ give the temperatures at the boundaries of each layer in the mantle. When we have a completely molten mantle, we only consider $T_{1}$ and $T_{2}$. $T_{1}$ is the temperature just below the surface while $T_{2}$ is the temperature at the bottom of the magma ocean. Once it cools below the liquidus, we consider $T_{3}$ and $T_{4}$ in addition to $T_{1}$ and $T_{2}$. $T_{3}$ and $T_{4}$ are the temperatures at the top and bottom of the mush layer. When the temperature goes below the solidus, we include $T_{5}$ and $T_{6}$, the temperatures at the top and bottom of the solid layer respectively. The average of $T_{2}$ and $T_{3}$ is $T_{i}$ while the average of $T_{4}$ and $T_{5}$ is $T_{i2}$. $T_{c}$ gives the temperature at the CMB. The differences between T$_3$ and T$_2$, T$_4$ and T$_5$, and T$_6$ and T$_c$, are due to the presence of the thermal boundary layers. 

The parameters $r_{1}$ to $r_{5}$ give the radial dimensions of the mantle layers at a given time. The terms $r_{2}$ and $r_{3}$ denotes the boundary lines just above and just below the interface at which the magma transitions to mush, and $r_{4}$ and $r_{5}$ are the boundary lines above and below the interface where the mush transitions to solid rock. The transition from magma to mush takes place at $r_{i}$ and is the average of $r_{2}$ and $r_{3}$. Similarly, $r_{i2}$ is the transition point from mush to rock and is the average of $r_{4}$ and $r_{5}$. To find these points we set the thermal profile to the liquidus and solidus. For the liquidus $T_{liq}$ and solidus $T_{sol}$ we used expressions derived from the high pressure experiments of \cite{Fiq2010} and \cite{zhang1994}. 

\begin{equation}
\label{Eqn1}
T_{liq} = 2000 \left(0.1169 \left(\frac{P}{GPa}\right) + 1\right)^{0.32726}
\end{equation}

\begin{equation}
\label{Eqn2}
T_{sol} = 1674 \left(0.0971 \left(\frac{P}{GPa}\right) + 1\right)^{0.35175}.
\end{equation}

We assumed the interior is fully convective and therefore the layers have adiabatic temperature profiles. We start simulations from a fully molten mantle, so $T_{c}$ can be extrapolated from $T_{s}$ by following an adiabatic geotherm. The model begins its run at the point where the geotherm of the fully molten mantle begins to dip below the liquidus curve. Following the approach of \cite{Lab2003} and \cite{Lab2015}, the relationship between $T_{1}$ and $T_{2}$ is 

\begin{equation}
\label{Eqn3}
T_{2} = T_{1} \exp \left(\frac{R_{p} - R_{c}}{D_{MO}}\right),
\end{equation}

\noindent where $D_{MO}$ is the temperature scale height of the magma ocean. Once mush and solid layer begin to form we determine $T_{4}$ and $T_{6}$ with 

\begin{equation}
\label{Eqn4}
T_{4} = T_{3} \exp \left(\frac{r_{3} - r_{4}}{D_{MS}}\right)
\end{equation}

\begin{equation}
\label{Eqn5}
T_{6} = T_{5} \exp \left(\frac{r_{5} - r_{6}}{D_{S}}\right),
\end{equation}

\noindent where $D_{MS}$ and $D_{S}$ are the temperature scale heights of the mush and rock respectively. 

Energy conservation in each layer gives the expressions describing the evolution of the temperature in each layer. These temperatures depend on the difference in heat flux entering and leaving the layer. For the magma ocean we evolve the temperatures $T_{s}$ and $T_{1}$. We used the Stefan-Boltzmann law to calculate the flux ($F$) radiated from the surface using the difference in incident stellar flux ($T_{eq}$) and the surface temperature. 

\begin{equation}
\label{Eqn6}
F =  \sigma \left(T_{s}^{4} - T_{eq}^{4}\right)
\end{equation}

\noindent Then to evolve $T_{s}$ we used the equation 

\begin{equation}
\label{Eqn7}
\frac{dT_{s}}{dt} = \frac{-F + Q_{MO} A_{MO}}{V_{MO} C \rho}
\end{equation}

\noindent The terms $Q$ and $A$ denote the heat flux and surface area for a given layer. $Q_{MO} A_{MO}$ is the heat lost by the magma ocean. $A_{MO}$ is $4 \pi R_{p}^{2}$ and $A_{MS}$ is $4 \pi r_{i}^{2}$. $V$, $C$ and $\rho$ denote the volume, specific heat capacity and density of a given layer of the mantle. For simplicity we use constant values for $C$ and $\rho$ which are given in Table \ref{tab1}. For the magma ocean, $V_{MO}$ is calculated as $\frac{4}{3} \pi (R_{p}^{3} - r_{i}^{3})$. For the evolution of $T_{1}$ we used the equation

\begin{equation}
\label{Eqn8}
\frac{dT_{1}}{dt} = \frac{-Q_{MO} A_{MO} + Q_{MS} A_{MS} + \frac{1}{2} L \rho A_{MS} \frac{dr_{i1}}{dt}}{V_{MO} C \rho}
\end{equation}

\noindent where $Q_{MS} A_{MS}$ is the heat flux entering the ocean from the mush below. $L$ is the latent heat released into the magma ocean due to the solidification from liquid to mushy state, i.e., 50 \% of melt. $\frac{dr_{i1}}{dt}$ gives the rate at which the mush front grows. We use similar energy conservation equations for the magma mush, solid rock and the CMB respectively denoted as 

\begin{equation}
\label{Eqn9}
\frac{dT_{3}}{dt} = \frac{-Q_{MS} A_{MS} + Q_{S} A_{S} + \frac{1}{2} L \rho A_{S} \frac{dr_{i2}}{dt}}{V_{MS} C \rho}
\end{equation}

\begin{equation}
\label{Eqn10}
\frac{dT_{5}}{dt} = \frac{-Q_{S} A_{S} + Q_{c} A_{c}}{V_{S} C \rho}
\end{equation}

\begin{equation}
\label{Eqn11}
\frac{dT_{c}}{dt} = \frac{-Q_{c} A_{c}}{V_{S} C \rho}.
\end{equation}

\noindent The terms $Q_{MS}$, $Q_{S}$ and $Q_{c}$ are the heat flux leaving the mush, solid rock and CMB respectively. The volumes of the mush and solid layers are calculated as $V_{MS} = \frac{4}{3} \pi (r_{i}^{3} - r_{i2}^{3})$ and $V_{S} = \frac{4}{3} \pi (r_{i2}^{3} - R_{c}^{3})$. 

We next assume that the heat flux leaving each layer is proportional to the temperature gradient across that layer. The expression used for the magma ocean is

\begin{equation}
\label{Eqn12}
Q_{MO} = 0.22 \kappa \left(\frac{T_{1} - T_{s}}{R_{p} - r_{i}}\right) \left(Ra_{MO}\right)^{\frac{2}{7}} Pr^{\frac{-1}{7}}
\end{equation}

\noindent where $\kappa$ is the thermal conductivity of the magma ocean, $Ra_{MO}$ the Rayleigh number and $Pr$ its Prandtl number. We look at the convective efficiency between $R_{p}$ and $r_{i}$ to compute the Rayleigh number of the magma ocean. We used the temperature difference between the surface and thermal interface ($T_{i}$ and $T_{s}$) corrected by the adiabatic gradient. We get the sum of this difference when calculating the Rayleigh number:

\begin{equation}
\label{Eqn13}
Ra_{MO} = \frac{ \alpha \rho g \left(T_{i} - T_{s} - (T_{2} - T_{1}) \right) \left(R_{p} - r_{i}\right)^{3}}{\eta_{MO} \kappa _{d}},
\end{equation}

\noindent where $\alpha$ is the thermal expansivity and $\kappa_{d}$ is the thermal diffusivity. We found $Pr$ with the equation

\begin{equation}
\label{Eqn14}
Pr = \frac{\eta_{k}}{\kappa_{d}}, 
\end{equation}

\noindent where $\eta_{k}$ is the kinematic viscosity. The heat flux through the mush layer, $Q_{MS}$, is controlled by the thermal properties of the mush. We used the temperature difference between the boundary layer at the upper portion of the mush and the ocean-mush interface to find the heat flux with 

\begin{equation}
\label{Eqn15}
Q_{MS} = 0.22 \kappa \left(\frac{T_{3} - T_{i}}{r_{i} - R_{c}}\right) \left(Ra_{MS}\right)^{\frac{1}{3}}
\end{equation}

\begin{equation}
\label{Eqn16}
Q_{MS} = 0.22 \kappa \left(\frac{T_{3} - T_{i}}{r_{i} - r_{i2}}\right) \left(Ra_{MS}\right)^{\frac{1}{3}}
\end{equation}

\noindent If we have the CMB right below the magma mush, we use Equation \ref{Eqn15} or if it is a solidification front right below, we use Equation \ref{Eqn16}. Since we are looking at layers of mush and solid rock, in which inertial forces are negligible. The Rayleigh number for the mush is defined as 

\begin{equation}
\label{Eqn17}
Ra_{MS} = \frac{ \alpha \rho g (T_{c} - T_{i} + T_{3} - T_{4}) (r_{i} - R_{c})^{3}}{\eta_{MS} \kappa _{d}}
\end{equation}

\begin{equation}
\label{Eqn18}
Ra_{MS} = \frac{ \alpha \rho g (T_{i2} - T_{i} + T_{3} - T_{4}) (r_{i} - r_{i2})^{3}}{\eta_{MS} \kappa _{d}}
\end{equation}

\noindent where we use Equation \ref{Eqn17} if there is no solid layer and Equation \ref{Eqn18} if the solid layer is below the mush. The equations of flux and Rayleigh number for the layer of solid rock are

\begin{equation}
\label{Eqn19}
Q_{S} = 0.22 \kappa \left(\frac{T_{5} - T_{i2}}{r_{i2} - R_{c}}\right) \left(Ra_{S}\right)^{\frac{1}{3}}
\end{equation}

\begin{equation}
\label{Eqn20}
Ra_{S} = \frac{ \alpha \rho g (T_{c} - T_{i2} + T_{5} - T_{6}) (r_{i2} - R_{c})^{3}}{\eta_{S} \kappa _{d}}
\end{equation}

\noindent Finally for the CMB we use the same Rayleigh number as what is given in Equation \ref{Eqn20}, while the heat flux can be evaluated with the expression 

\begin{equation}
\label{Eqn21}
Q_{c} = 0.22 \kappa \frac{(T_{c} - T_{4})}{r_{i} - R_{c}} (Ra_{s})^{\frac{1}{3}}
\end{equation}

\begin{equation}
\label{Eqn22}
Q_{c} = 0.22 \kappa \frac{(T_{c} - T_{6})}{r_{i2} - R_{c}} (Ra_{S})^{\frac{1}{3}}
\end{equation}

\noindent where Equation \ref{Eqn21} and \ref{Eqn22} are if we have mush or solid rock above the CMB respectively.

\subsection{Horizontal heat transfer}  \label{sec2.2}

While the vertical convective currents are driven by the temperature differences between the surface and the CMB, we also get horizontal convective currents driven by the day-to-night temperature difference. In our model the day-side surface temperature is fixed at $T_{s} =$ 3000 K while the night-side has a time varying $T_{s}$. We assume that most of the horizontal heat transfer happens in the layers with the lowest viscosities. We add a sink/source term to the energy conservation equations of the magma oceans on the day and night sides. We use a thermal Rayleigh number $Ra_{H}$ for horizontal convection from \cite{Hugh2008}. $Ra_{H}$ differs in that we use the difference in $T_{s}$ between the day-side and night-side for the temperature contrast and we use half the planetary circumference as the length. We can then have the expression   

\begin{equation}
\label{Eqn23}
Ra_{H} = \frac{ \alpha \rho g (T_{sd} - T_{sn}) (\pi R_{p})^{3}}{\eta_{ns} \kappa _{d}}
\end{equation}

\noindent where $T_{sd}$ and $T_{sn}$ are the day-side and night-side surface temperatures respectively. The parameter $\eta_{ns}$ is the magma viscosity, while $\rho_{ns}$ is the density of that same layer. We then evaluate the horizontal heat flux $Q_{H}$ with the expression 

\begin{equation}
\label{Eqn24}
Q_{H} = 0.22 \kappa \frac{(T_{sd} - T_{sn})}{\pi R_{p}} (Ra_{H})^{\alpha} .
\end{equation}

\noindent We set the value of $\alpha$ to $\frac{2}{7}$ under the assumption we have turbulent convection in the magma \citep{Hugh2008}. By adjusting $\eta_{ns}$, we tested different quantities of heat flux flowing to the night-side. When $\eta_{ns}$ increases, the magnitude of horizontal heat flow will gets weaker. To account for horizontal heat flow from the day-side we added $Q_{H}$ to Equation \ref{Eqn7} as applied to the day-side magma ocean. For the night-side we reduced $Q_{H}$ from Equation \ref{Eqn7} as applied to the night-side magma ocean. This gave new expressions for the day-side and night side evolution of the surface temperature.

\begin{equation}
\label{Eqn25}
\frac{dT_{1d}}{dt} = \frac{-Q_{MO} A_{MO} + Q_{MS} A_{MS} + \frac{1}{2} L \rho A_{MS} \frac{dr_{id}}{dt} - Q_{H}}{V_{MO} C \rho}
\end{equation}

\begin{equation}
\label{Eqn26}
\frac{dT_{1n}}{dt} = \frac{-Q_{MO} A_{MO} + Q_{MS} A_{MS} + \frac{1}{2} L \rho A_{MS} \frac{dr_{in}}{dt} + Q_{H}}{V_{MO} C \rho}
\end{equation}

\subsection{Tidal heating}   \label{sec2.3}

Because of the proximity of lava planets to their host stats, it is likely that they experience strong tidal forces assuming they are not tidally locked \citep{Leger2011}. To include the effect of tidal heating $Q_{T}$, we assumed that most of the tidal dissipation happens in the mush rather than in the magma or solid rock \citep{Kerv2021}. Therefore we scaled the magnitude of tidal dissipation with the mass of the mush at a given time. To find the amount of tidal energy dissipation per unit mass, we used the data from \cite{Ty2015}. Their work simulates the tidal heating taking place within the magma ocean of Io. They concluded that the solid part of Io could have tidal dissipation in the amount of 2.25 Wm$^{-2}$. Converting the energy units from Wm$^{-2}$ to total energy dissipation, we get $\approx 10^{14}$ Watts within Io. Dividing the total energy dissipation by the mass of Io and then scaling that value by the mass of mush in our model planets, we find $Q_{T}$. 

\begin{equation}
\label{Eqn27}
Q_{T} = f \times \frac{10^{14}}{M_{Io}} \times M_{mush} .
\end{equation}

\noindent The value of $Q_{T}$ can be increased or decreased by the scaling factor $f$. We add $Q_{T}$ to Equation \ref{Eqn7} along with $Q_{H}$.

\begin{equation}
\label{Eqn28}
\frac{dT_{1}}{dt} = \frac{-Q_{MO} A_{MO} + Q_{MS} A_{MS} + L \rho A_{MS} \frac{dr_{i}}{dt} + Q_{H} + Q_{T}}{V_{MO} C \rho}
\end{equation}

\noindent To investigate if the quantities of tidal heating needed to keep a molten night-side are possible, we calculated the relationship between orbital eccentricity ($e$) and tidal dissipation ($Q_{tidal}$). We used the tidal power equation from \cite{Dris2015}. 

\begin{equation}
\label{Eqn29}
Q_{tidal} = -\frac{21}{2} Im(k2) G^{\frac{3}{2}} M_{s}^{\frac{5}{2}} R_{p}^{5} \frac{e^{2}}{a^{15}}
\end{equation}

\noindent The parameters $G$, $M_{s}$, $a$ and $R_{p}$ are the gravitational constant, stellar mass, semi-major axis and planetary radius respectively. $Im(k2)$ is the imaginary component of the Love number k2, and was given a value of $3 \times 10^{-3}$ for an Earth-like composition. We used this value of $Im(k2)$ despite it being for a more viscous medium than mush, to investigate what quantities of $Q_{T}$ are possible with it. 

The parameters $Q_{H}$ and $Q_{T}$ can be turned on or off to simulate different combinations of heating added to the night-side. The parameter $T_{eq}$ in Equation \ref{Eqn8} was fixed at 3000 K for the day-side, but given a value of zero for the night-side since there is no incoming stellar flux. The parameter $T_{c}$ was given the same value for both the day and night sides. The two hemispheres of the planet were evolved while exchanging heat from the day-side to the night-side only in the top-most layer. The simulations were run until no more changes were seen in the internal structure for each hemisphere of a given planet.

\section{Results}  \label{sec3}

We applied our model to lava planets with radii of 1.0$R_{\oplus}$ and 1.5$R_{\oplus}$, the former for bench-marking with terrestrial models and the latter to simulate K2-141b. We evaluated each radius for a core mass fraction (CMF) of 32 per cent and 70 per cent. The CMF of 32 per cent mimics an Earth-like interior while the CMF of 70 per cent is for a Mercury-like interior. Our models showed that the latent heat released by the solidification of the magma ocean was smaller in magnitude by a factor of at least 100 at their highest value, compared to the radiated heat flux and the convective heat flux. This was because the rate of growth of mush and solid was too slow to have an effect on the latent heat quantity. The latent heat shows a steady decline in magnitude with a peak around $2 \times 10^{18}$ Watts for a 1.0$R_{\oplus}$ planet. This is when the growth rate is at its maximum value of $10^{-5}$ ms$^{-1}$. The peak energy was even lower for 1.5$R_{\oplus}$ on account of their mush growth rates being lower. It never reached a level within two orders of magnitude of the energy lost by radiation or the energy contribution via day-night heating and tidal dissipation. Therefore we will not take the effect of Latent heat into account in the rest of this study.     

\subsection{Earth-size planets}    \label{sec3.1}

\begin{figure*}\centering
\includegraphics[width=0.48\textwidth]{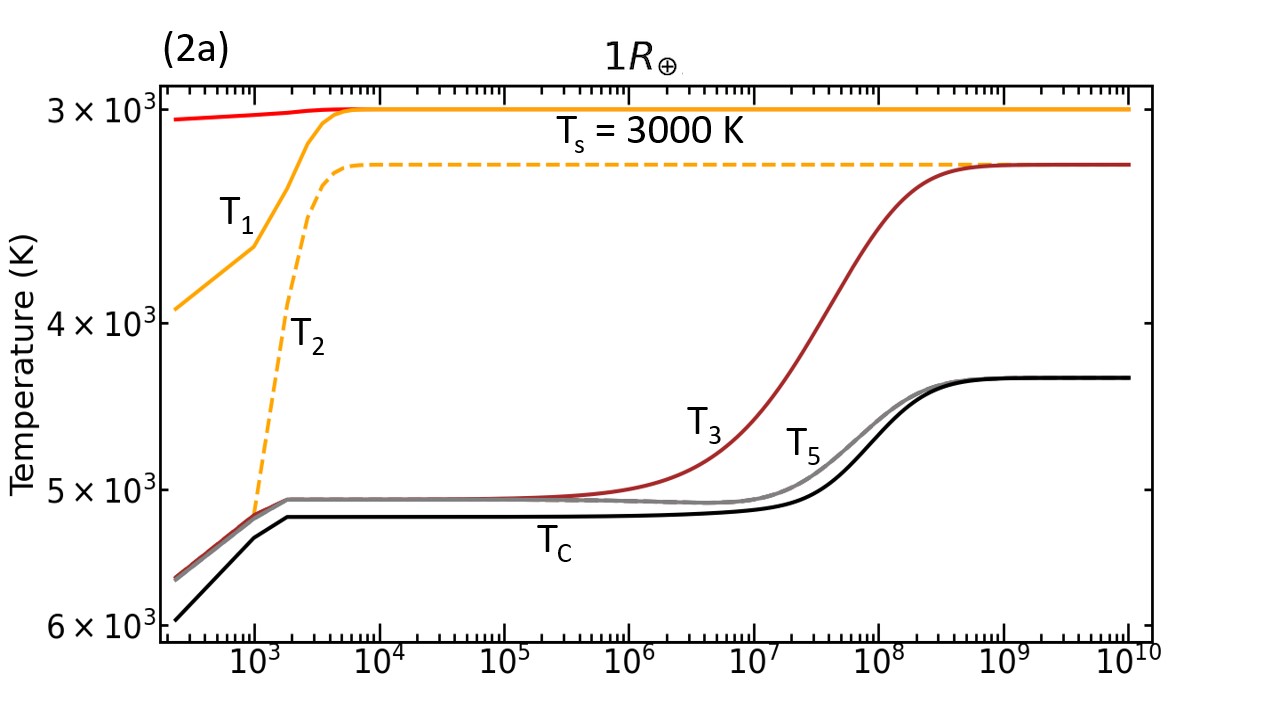}
\includegraphics[width=0.48\textwidth]{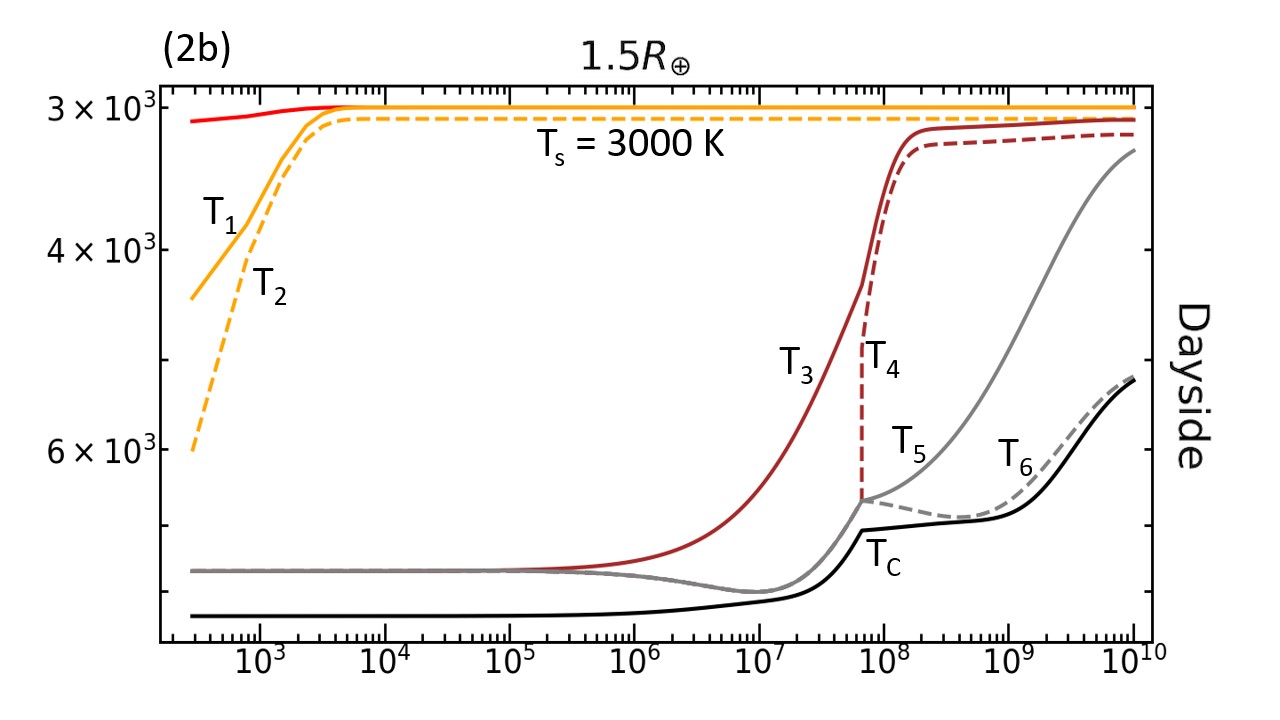}
\includegraphics[width=0.48\textwidth]{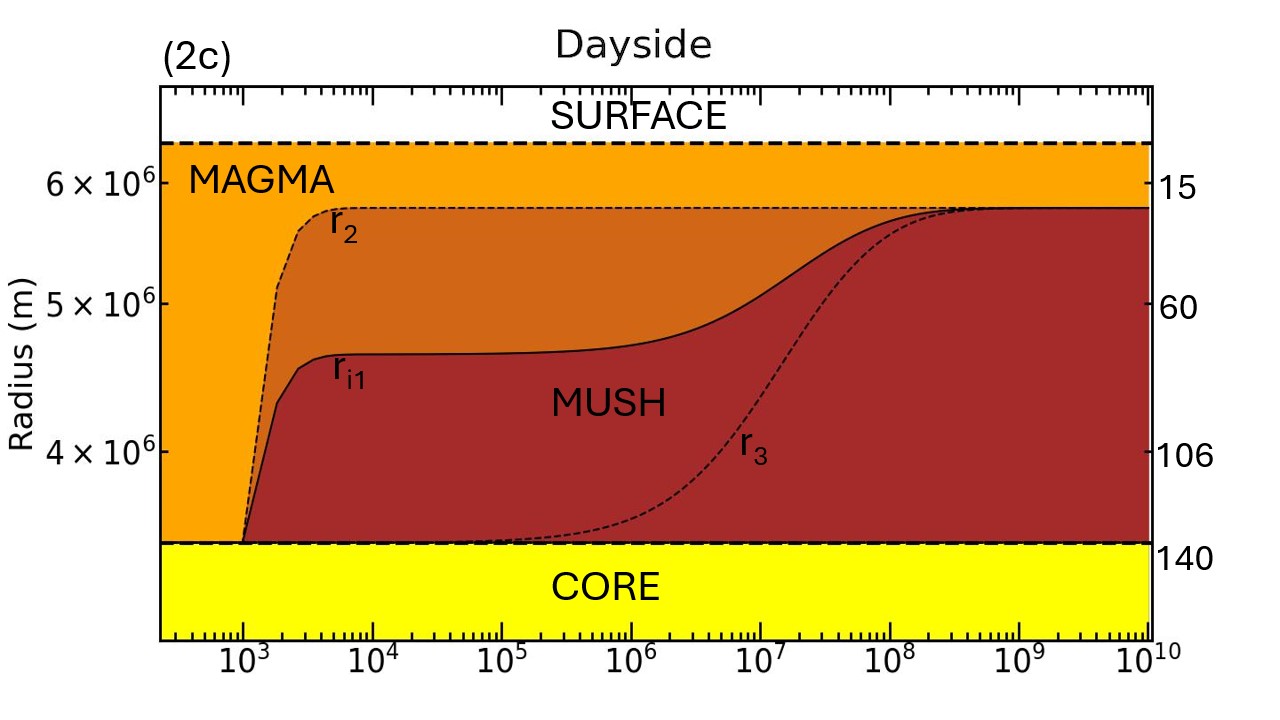}
\includegraphics[width=0.48\textwidth]{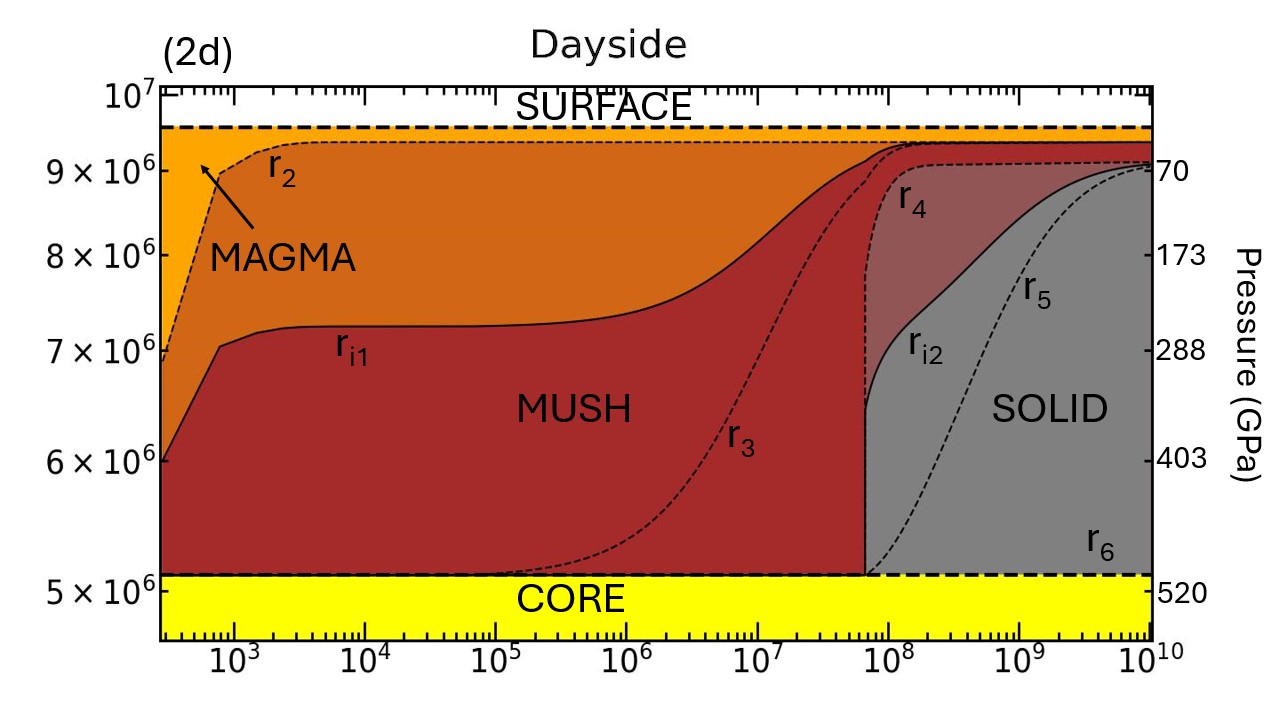}
\includegraphics[width=0.48\textwidth]{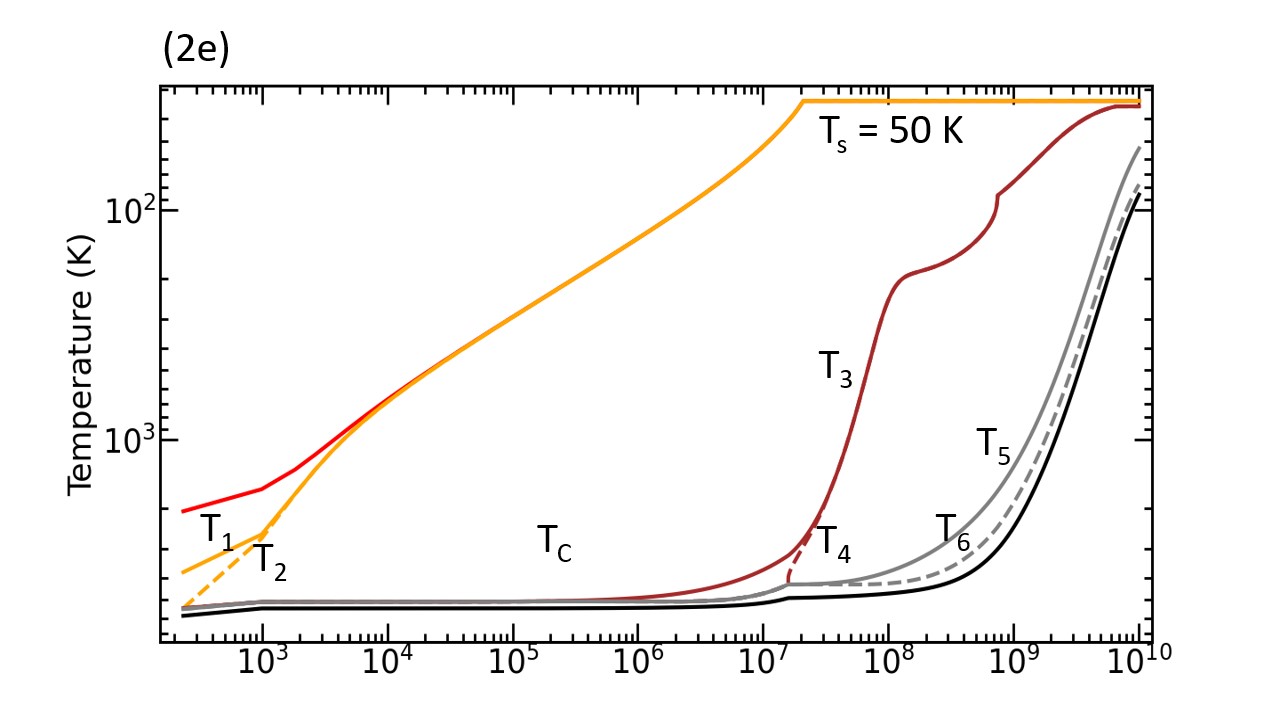}
\includegraphics[width=0.48\textwidth]{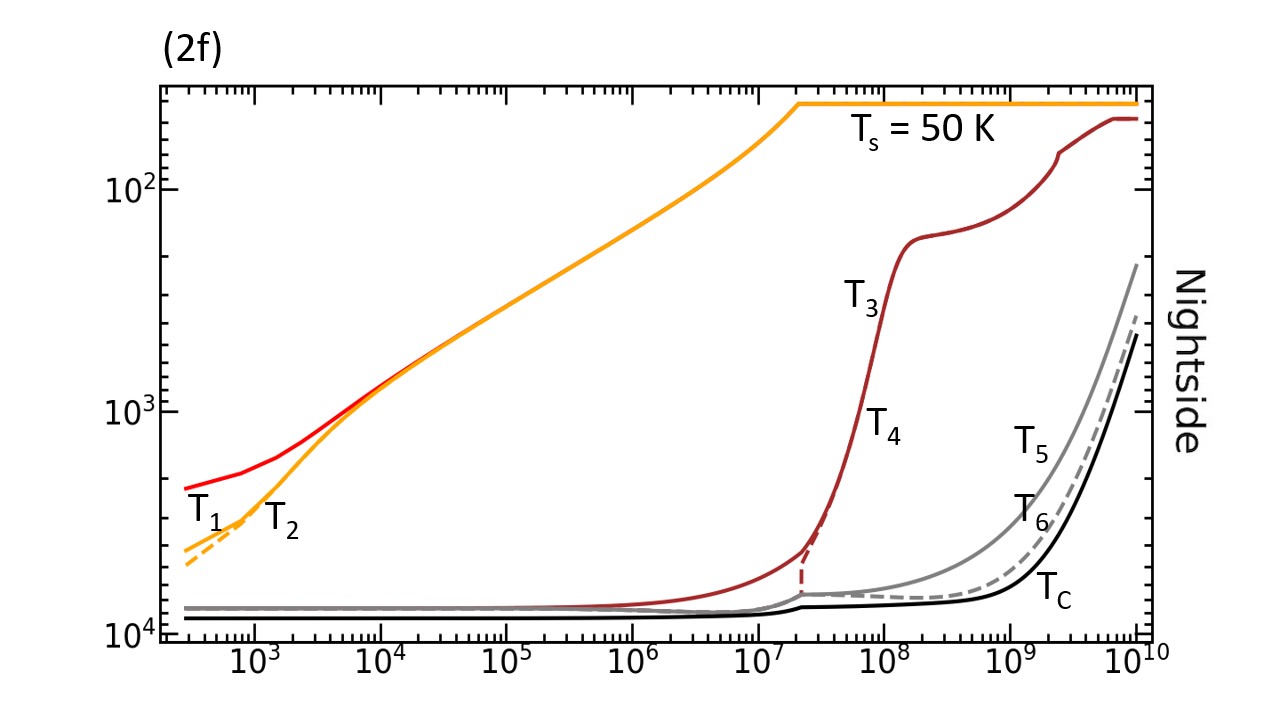}
\includegraphics[width=0.48\textwidth]{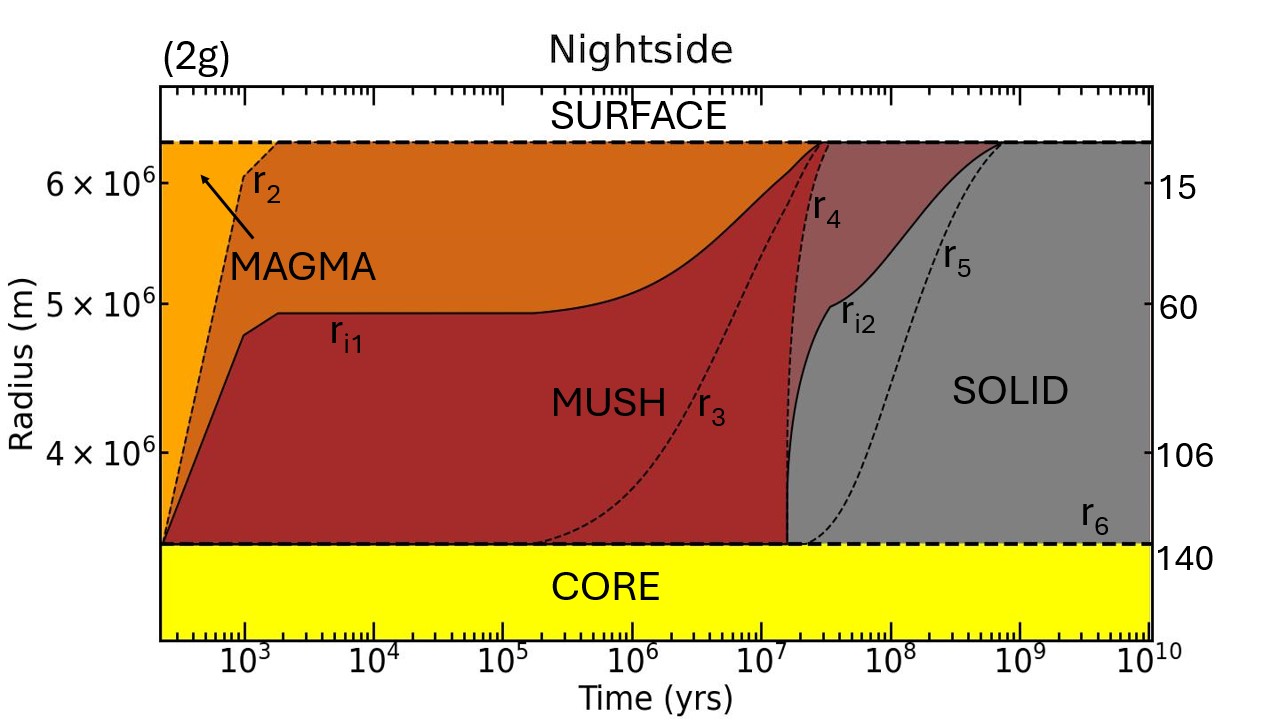}
\includegraphics[width=0.48\textwidth]{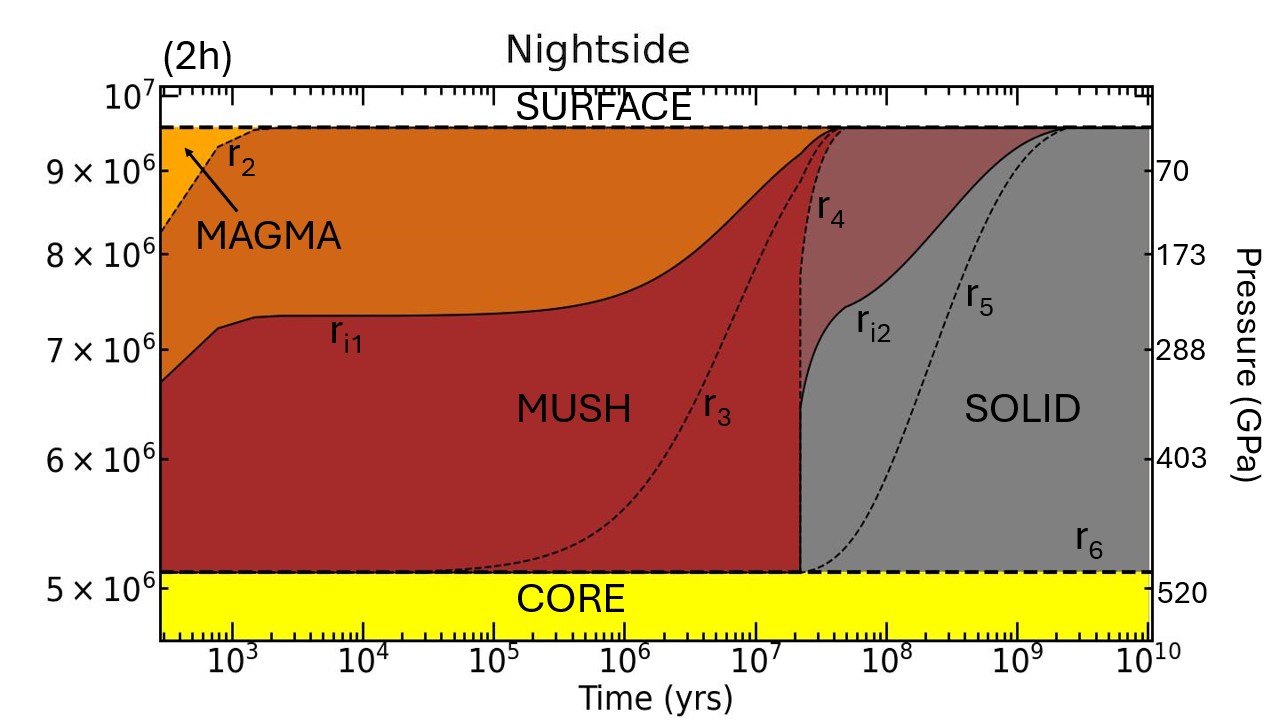}

\caption{The thermal and internal evolution of a 1.0$R_{\oplus}$ planet (left) and a 1.5$R_{\oplus}$ planet (right) with a core mass fraction of 32 per cent in the absence of day-to-night heat transport or tidal heating. The top four panels show the evolution of the day-side while the bottom four panels show the night-side evolution. The coloured lines show the temperatures at the boundaries shown in Figure \ref{fig1}. In panel 2a, the temperatures $T_{4}$ and $T_{6}$ are overlapping $T_{3}$ and $T_{5}$ respectively. The temperature increases downwards in these plots to ease comparison with the structure model. The coloured shading shows how the depth of the magma, mush and solid rock change in time. The sections where the colours have are translucent are the regions where phase transitions are happening (i.e. mush and solid crystals). The day-side of the 1.0$R_{\oplus}$ planet maintains a 550 km deep magma ocean in a steady state and the day-side of the 1.5$R_{\oplus}$ has a shallower magma ocean of about 200 km. The 1.5$R_{\oplus}$ planet has a shallow magma ocean and more solid rock in the day-side interior because of the higher gravity of the bigger planet. The night-side of both planets solidify completely within approximately $10^{9}$ years; the smaller planet does so faster due to its smaller thermal inertia.}
\label{fig2}
\end{figure*}

Figure \ref{fig2} shows the thermal and internal evolution of the day-side and night-side of a 1.0$R_{\oplus}$ planet with a CMF of 32 per cent. On the day-side, the surface reaches radiative equilibrium with the incoming stellar flux at a temperature of 3000 K. This is high enough to keep a stable layer of magma at the top of the mantle. The high level of flux leaving the magma ocean at the onset of cooling on the day-side reduces rapidly before reaching an equilibrium state with the flux from the magma mush. This only lasts for a brief time period before the flux from the mush reduces and itself reaches equilibrium with the flux from the newly formed solid rock at the bottom. On the night-side, the lower surface temperature and the absence of incoming flux leads to a higher magnitude of cooling. This results in the total solidification of the night-side after about 500 million years. Figure \ref{fig2} shows the evolution of the two sides of the planet without horizontal heat transfer from the day-side to the night-side. The rapid cooling rate on the night-side would mean there is a large temperature contrast between the magma oceans of the two sides at a given time. Therefore heat transfer from the hotter day-side to the night-side is highly probable.

\begin{figure*}\centering
\includegraphics[width=0.48\textwidth]{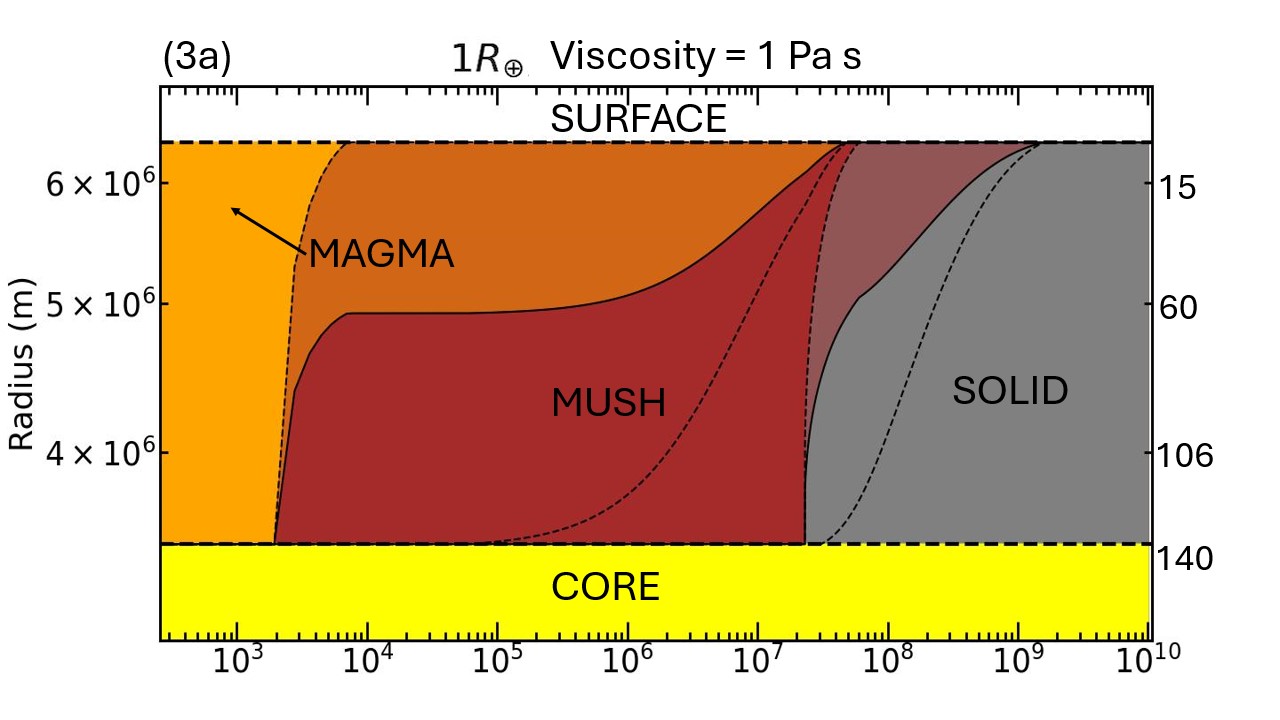}
\includegraphics[width=0.48\textwidth]{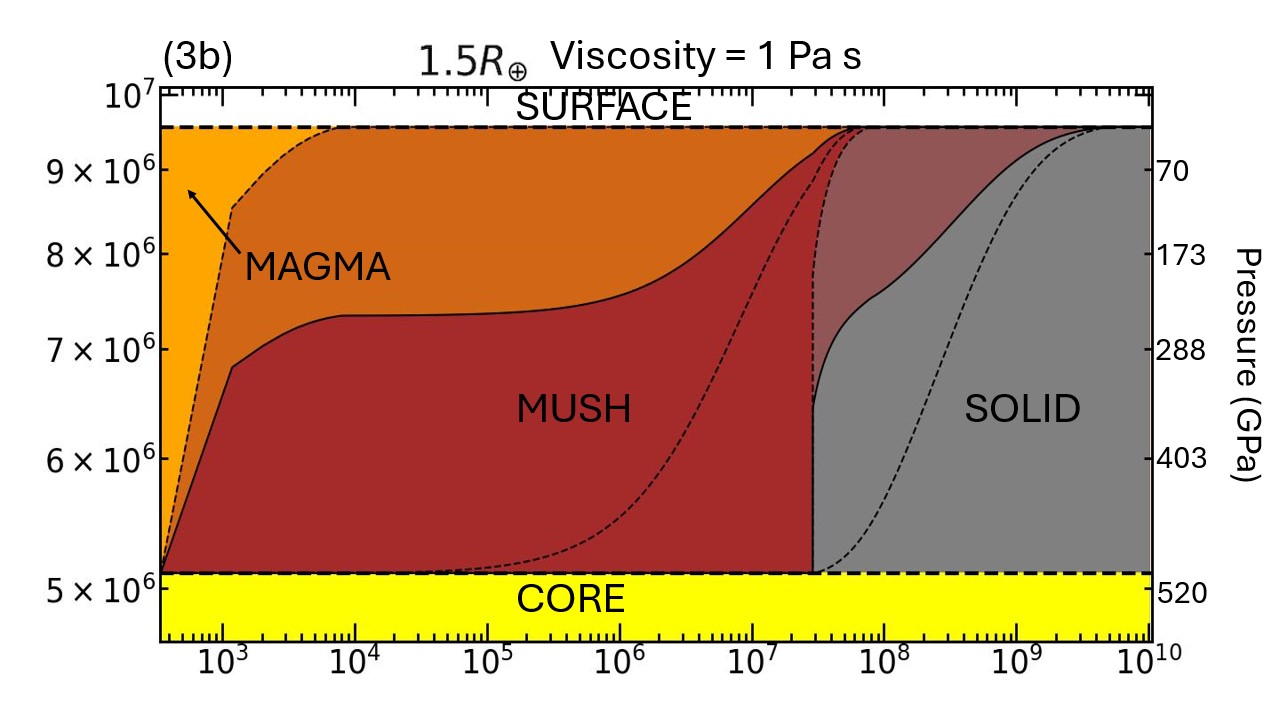}
\includegraphics[width=0.48\textwidth]{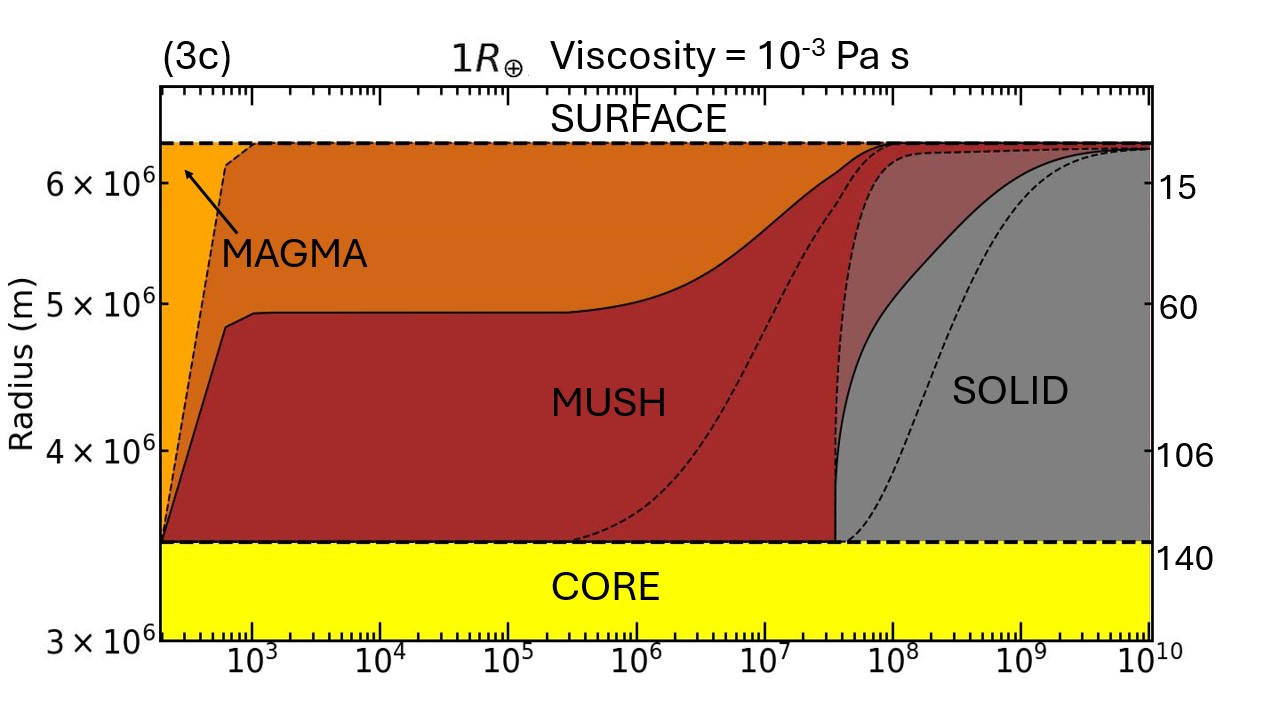}
\includegraphics[width=0.48\textwidth]{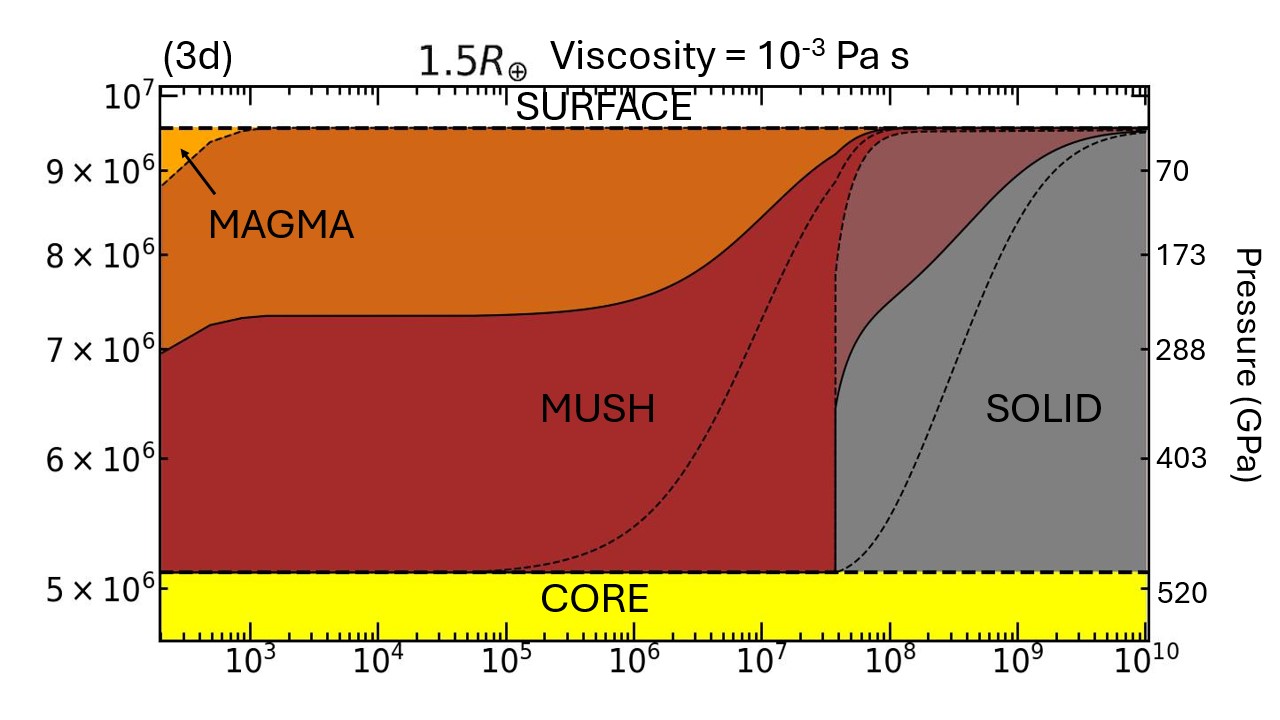}
\includegraphics[width=0.48\textwidth]{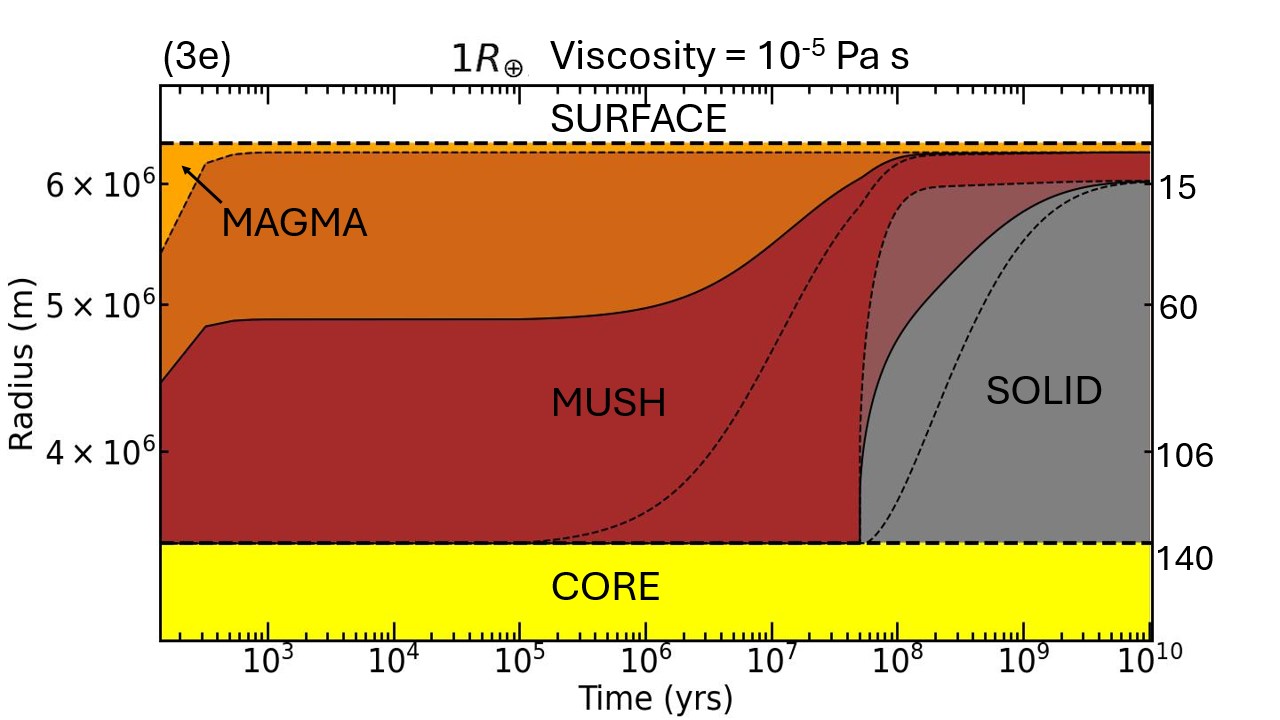}
\includegraphics[width=0.48\textwidth]{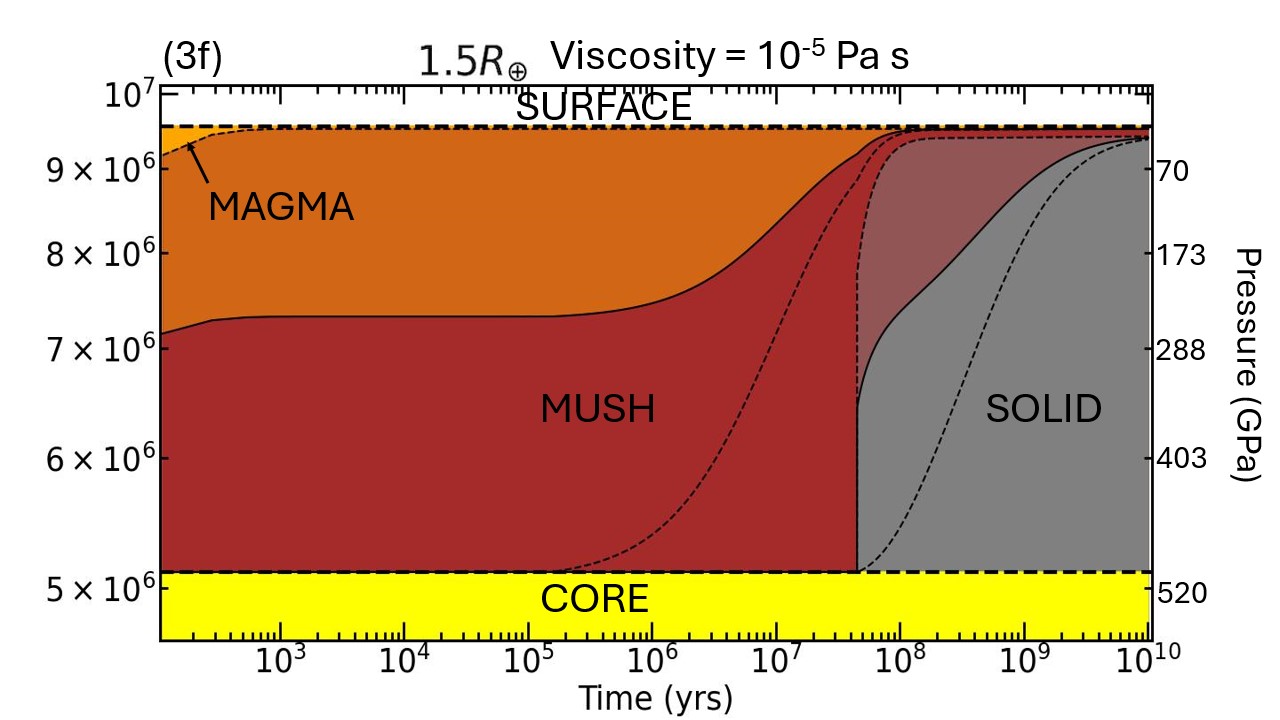}

\caption{The interior evolution of the night-side of 1.0$R_{\oplus}$ (left) and 1.5$R_{\oplus}$ (right) planets with a core mass fraction of 32 per cent} and the inclusion of horizontal heat transport $Q_{H}$. Panels (a) and (b) show the evolution at a magma viscosity of $\eta = 1$ Pa s, (c) and (d) for $\eta = 10^{-3}$ Pa s and (e) and (f) for $\eta = 10^{-5}$ Pa s. In (a) and (b) we can see the night-side fully solidifying while (c) and (d) show a scenario where the night-side surface remains mushy. Panels (e) and (f) show that the night-side surface can be kept molten for both Earth-sized and Super-Earth lava planets if the magma viscosity is sufficiently low.
\label{fig3}
\end{figure*}

We re-ran the simulations with the inclusion of horizontal heat transport $Q_{H}$ in the model. From Equation \ref{Eqn23} it can be shown that the viscosity of the magma controls $Ra_{H}$. The value of $\eta$ in Equation \ref{Eqn23} was the same for the day and night sides. The magma viscosity was varied from $1$ Pa s to $10^{-5}$ Pa s and we recorded the values of $T_{s}$ resulting from each. A viscosity of $10^{-5}$ Pa s is an extreme scenario that we tested in order to explore the conditions under which a molten night-side surface could be maintained. In Figure \ref{fig3} we see how the state of the surface changes with $Q_{H}$. If we have a minimum $Q_{H}$ of $2.3 \times 10^{20}$ Watts, we observe a molten night-side surface. The energy transferred between hemispheres change with $\eta$. If the viscosity is between 1 and $10^{-2}$ Pa s ($5 \times 10^{19} < Q_{H} < 8 \times 10^{19}$ Watts), the night-side completely solidifies after 2 Gyrs. The surface temperature fluctuates between 1000 K and 1500 K at these values. Reducing the viscosity to values between $10^{-2}$ and $10^{-4}$ Pa s ($8 \times 10^{19} < Q_{H} < 2.3 \times 10^{20}$ Watts) results in a mush surface as well as surface temperatures between 1500 K and 2000 K. At $10^{-4}$ Pa s we start to see a very thin (about 40 km) sustained magma ocean on the night-side. Theoretically there is enough energy at this particular viscosity to keep the night-side surface melted, which is consistent with the model result. However, the viscosity is expected to increase when temperature decreases and crystal fraction increases. Hence the sustenance of a magma ocean could be impeded if we include a temperature-dependant viscosity. This means the existence of a magma ocean at $\eta = 10^{-4}$ Pa s may or may not be possible. Once the viscosity goes below $10^{-4}$ Pa s, $Q_{H}$ is high enough to maintain a night-side magma ocean. At $10^{-5}$ ($3.5 \times 10^{20}$ Watts) Pa s we could see a magma ocean about 120 km deep, meaning that a very low magma viscosity is necessary to have a melted night-side surface if we only take horizontal heat transfer into account. Given that a magma viscosity of $10^{-5}$ Pa s is an extreme scenario, it is unlikely that horizontal day-night convection alone could sustain a fully molten night-side magma ocean.

\begin{figure*}\centering
\includegraphics[width=0.48\textwidth]{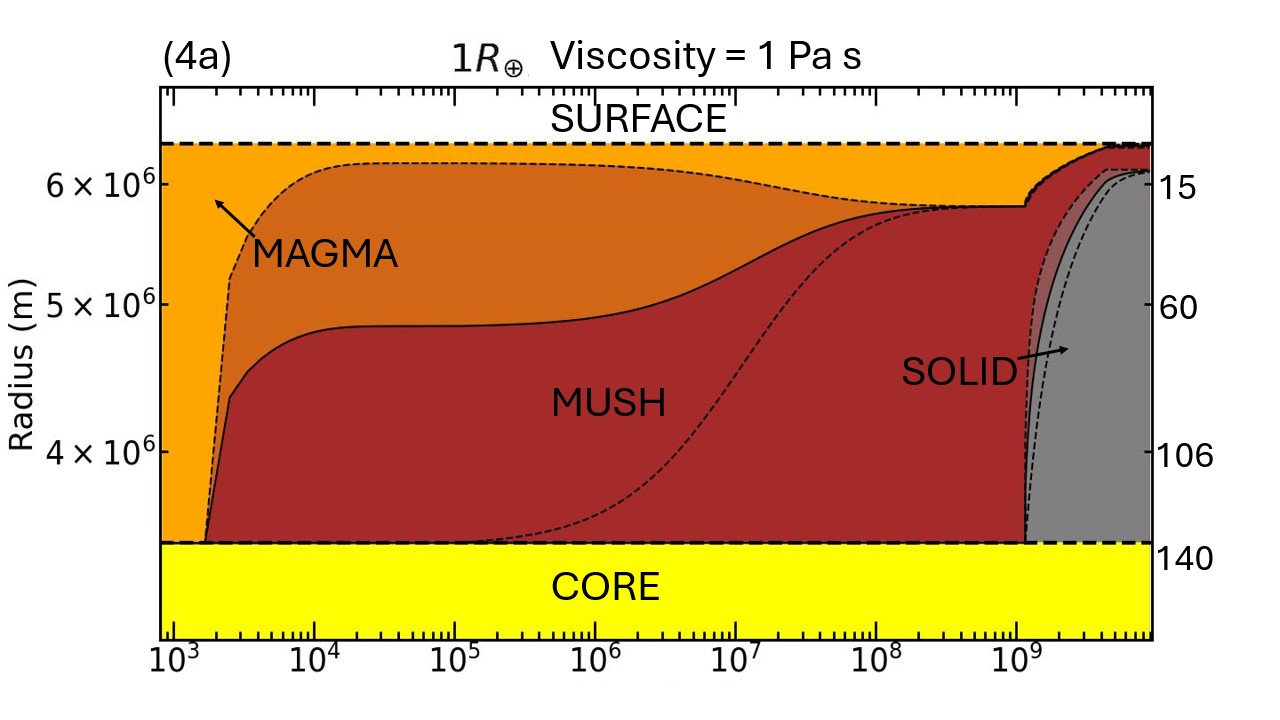}
\includegraphics[width=0.48\textwidth]{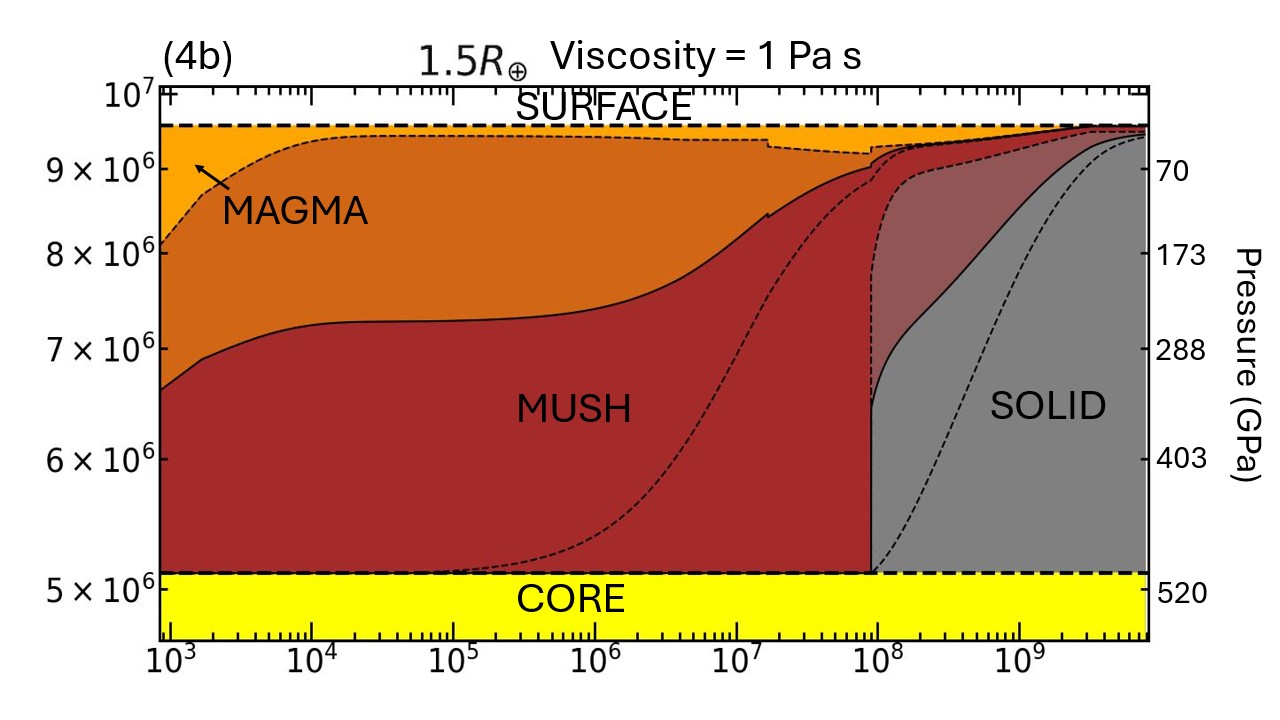}
\includegraphics[width=0.48\textwidth]{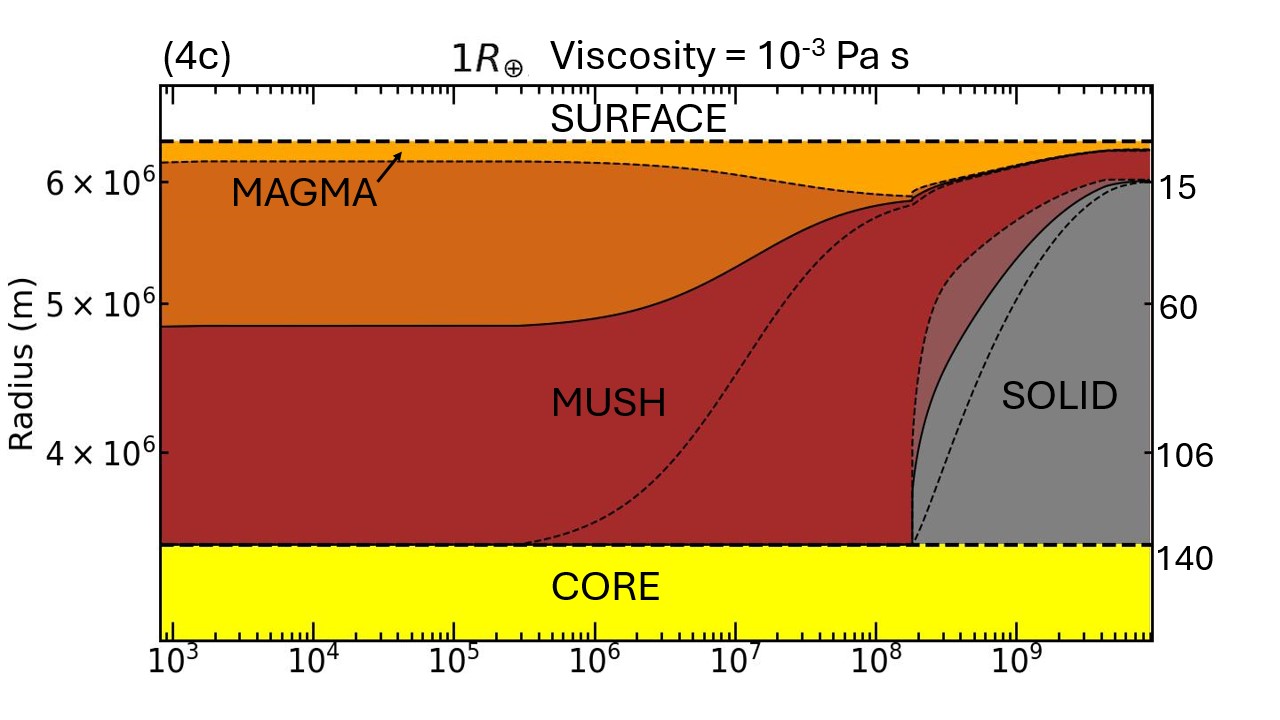}
\includegraphics[width=0.48\textwidth]{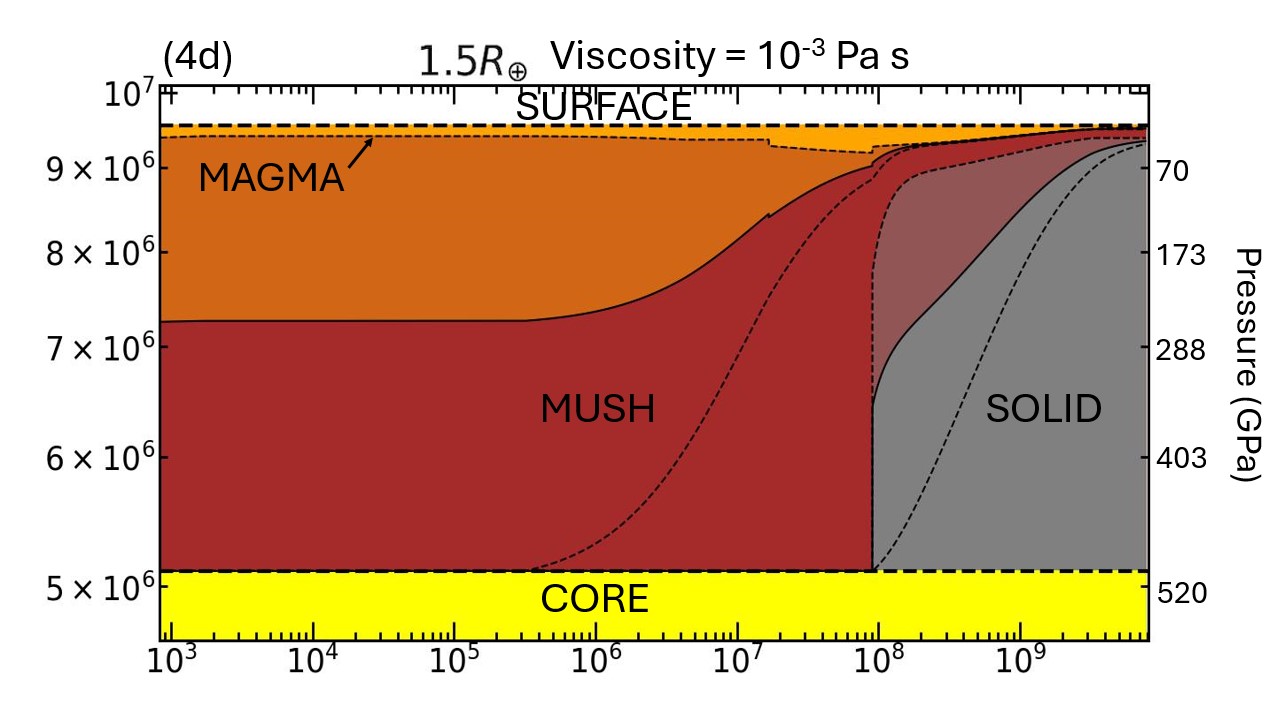}
\includegraphics[width=0.48\textwidth]{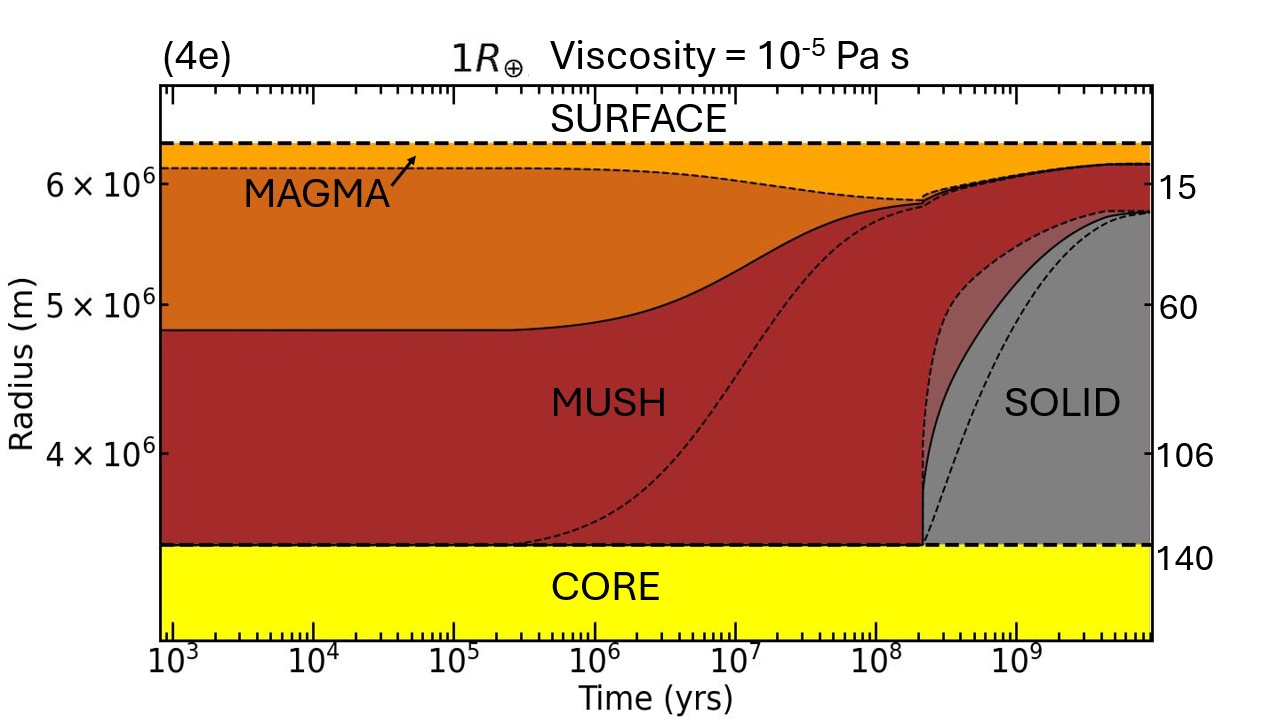}
\includegraphics[width=0.48\textwidth]{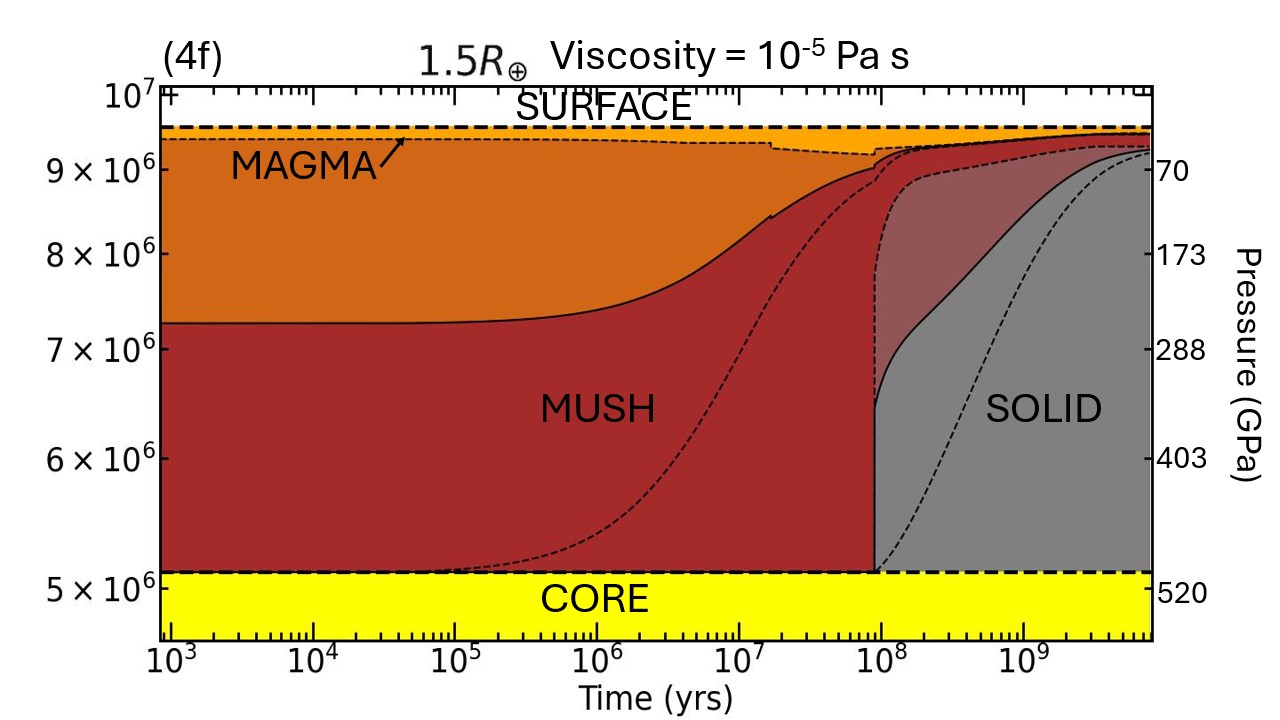}

\caption{Evolution of the interior with the inclusion of both day-night heat transport ($Q_{H}$) and tidal heating ($Q_{T}$). The wattage per kg of mush is $q_{T} = 7 \times 10^{-4}$ W/kg. The mush mass varies by a factor of $\sim 2$, peaking at $\leq 10^{8}$ years before dropping to $\sim$ zero after a few billion years. Tidal heating is greater for larger planets because of the greater mass of mush.}
\label{fig4}
\end{figure*}

We increased the CMF of the 1.0$R_{\oplus}$ planet to 70 per cent for the next set of simulations. As before, the initial simulations did not have any horizontal heat transport before being added later on. The day-side and night-side did not show much variation compared to the models with a CMF of 32 per cent. The only noticeable difference was that there was less solid rock in the deep interior of the day-side and the timescale for total solidification on the night-side happens about 600 million years earlier than it did for the 32 per cent CMF model. With the addition of horizontal heat transport, once again there was no discernible difference between the CMF = 32 and CMF = 70 per cent models. The value of $Q_{H}$ required to sustain a night-side magma ocean was the same as it was in the earlier models.

\begin{figure*}\centering
\includegraphics[width=0.48\textwidth]{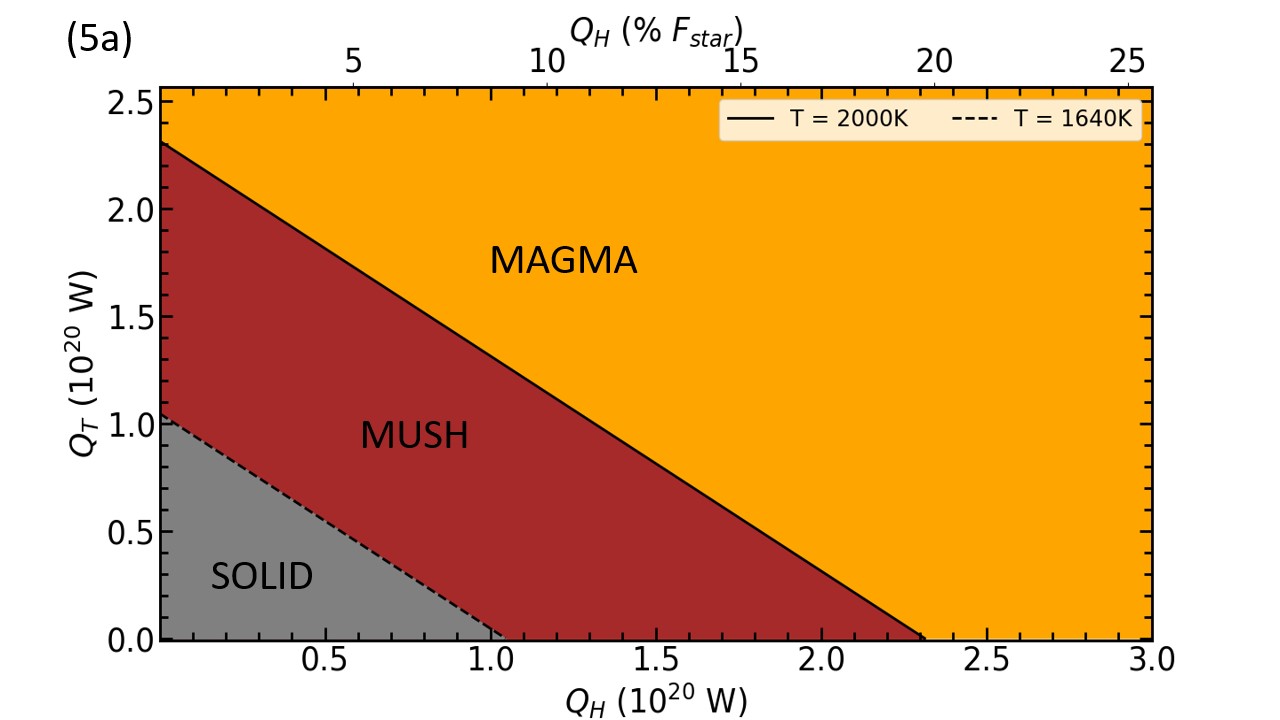}
\includegraphics[width=0.48\textwidth]{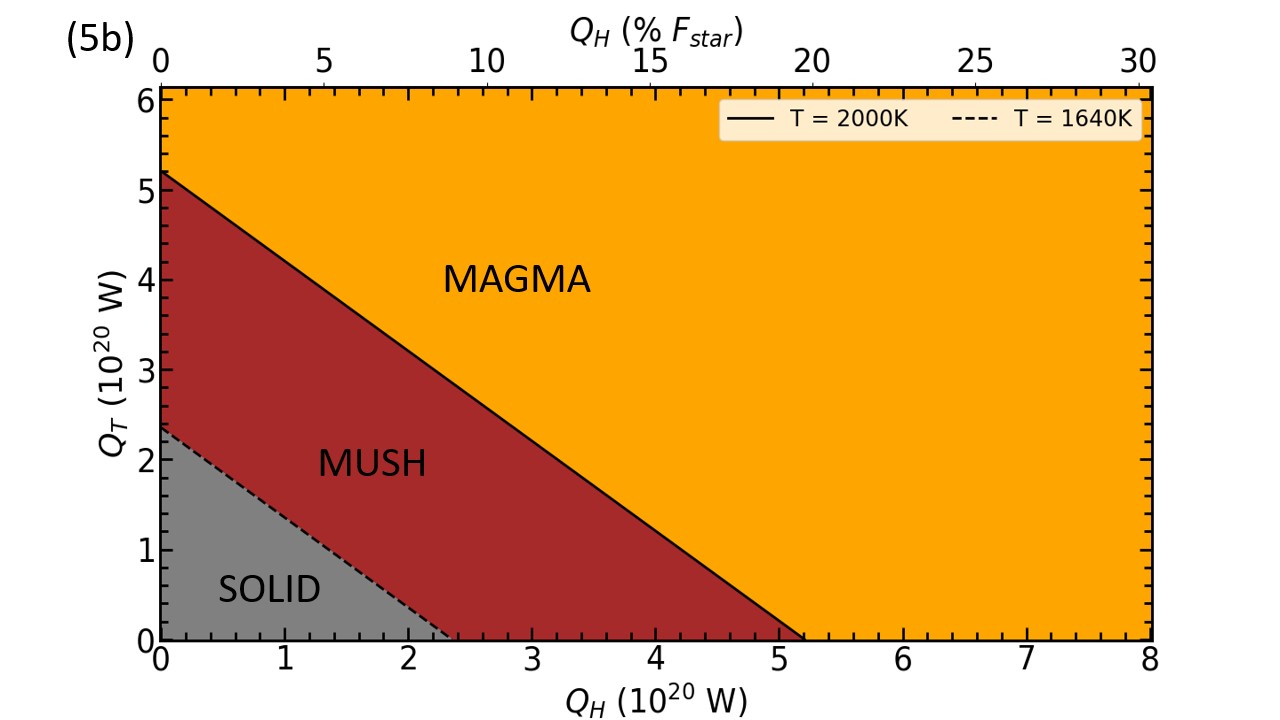}
\includegraphics[width=0.48\textwidth]{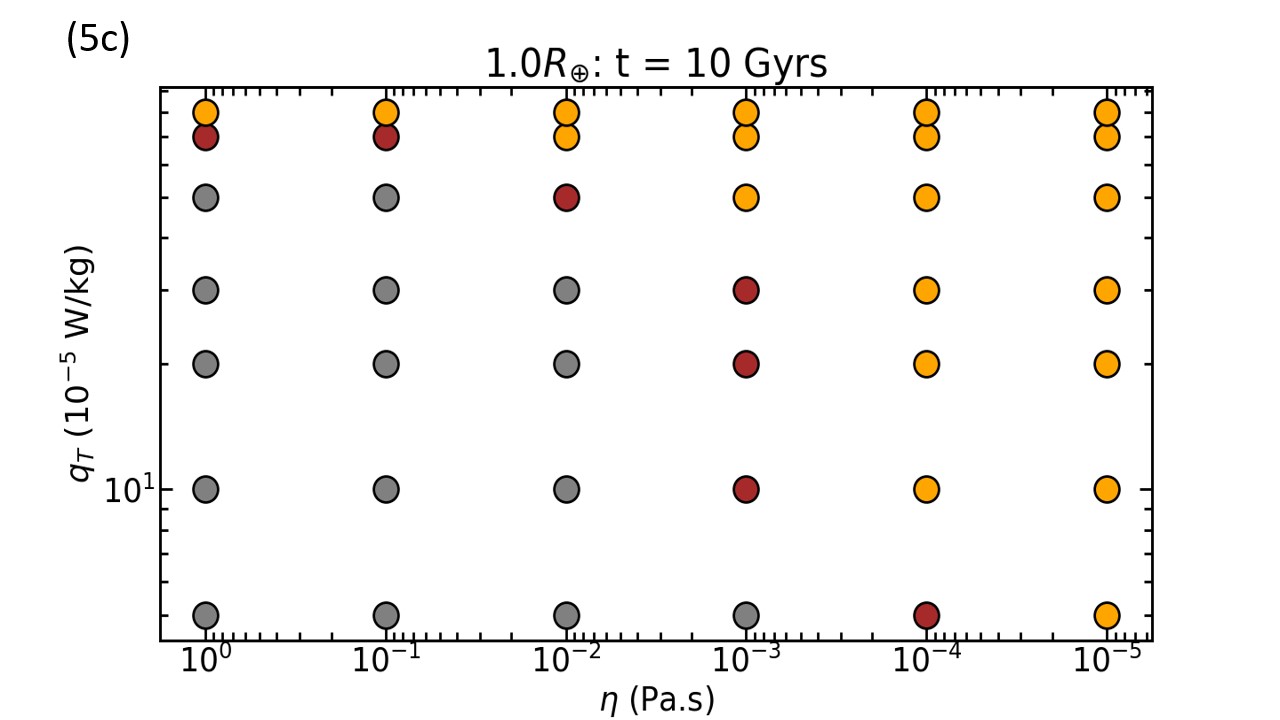}
\includegraphics[width=0.48\textwidth]{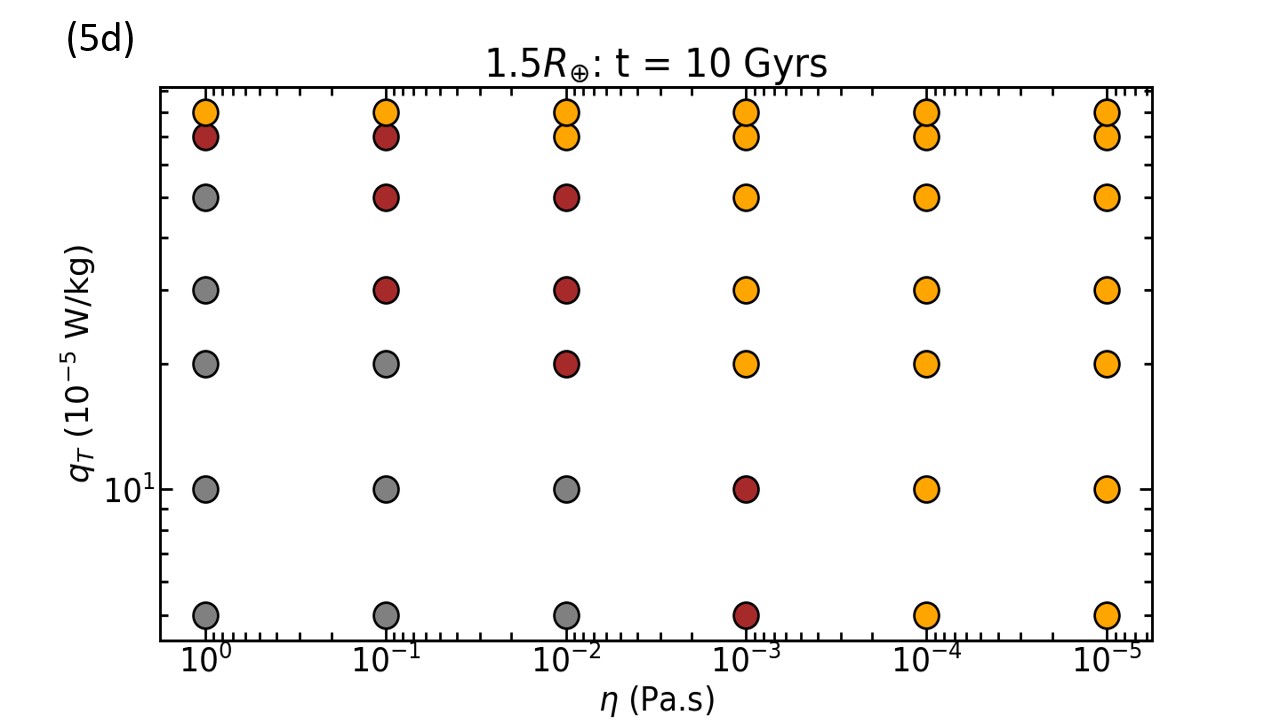}

\caption{Night-side surface end-state as a function of day-night heat transport ($Q_{H}$), and night-side tidal heating ($Q_{T}$). Panels 5a (1.0$R_{\oplus}$ planets) and 5b (1.5$R_{\oplus}$ planets) show the theoretical 2D parameter space based on the Stefan-Boltzmann formula for night-side cooling. At the top of each panel we show the horizontal heat transport $Q_{H}$ as a percentage of the instellation. Panels 5c and 5d show a grid of 42 numerical simulations using our thermal evolution model. These figures show the viscosity $\eta$ and specific tidal heating, $q_{T}$, used in each simulation. The colour of each datum gives the state of the night-side surface after running the simulation for 10 Gyrs. The upper and lower panels show qualitatively the same pattern.}
\label{fig5}
\end{figure*}

\subsection{Super-Earth's}   \label{sec3.2}

\begin{figure}\centering
\includegraphics[width=0.5\textwidth]{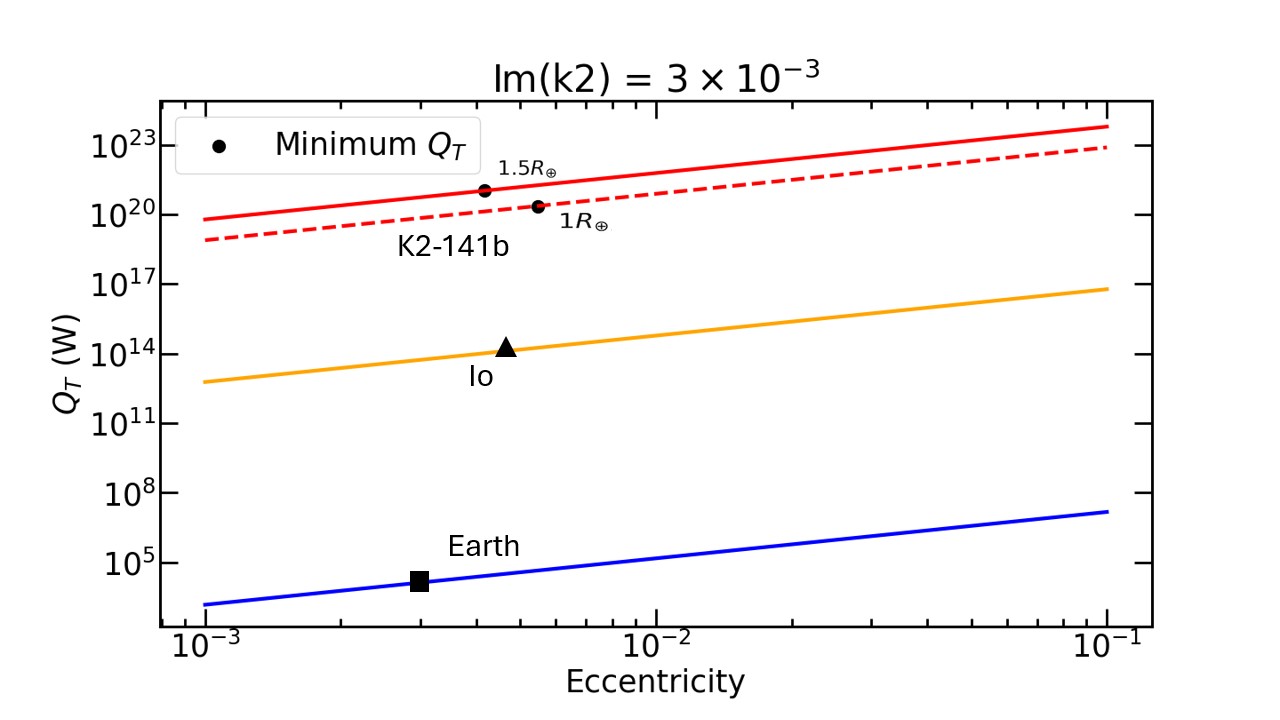}

\caption{A comparison of the tidal heating energy generated in Earth (blue), Io (orange) and K2-141 b (red) as a function of orbital eccentricity. For K2-141b, the solid line is for the planet's actual 1.5$R_{\oplus}$ radius, while the dashed line shows the tidal heating if the planet was Earth-sized. The calculations presume an Earth-like $\rm{Im}(k2) = 3 \times 10^{-3}$ for an Earth-like composition. The black dots show the minimal tidal energy required for the night-side to remain melted, in the absence of day-to-night transport (see top panels of Figure 5). The black triangle and square shows the average tidal energy dissipated inside Io and Earth respectively.}
\label{fig6}
\end{figure}

For the next set of models, the radius of the planet was increased to 1.5$R_{\oplus}$ to approximate a Super-Earth. In Figure \ref{fig2}, the right-side panels of 2b, 2d, 2f and 2h show the evolution of the internal structure of 1.5$R_{\oplus}$ model for the day-side and night-side. Compared to Earth-sized planets, there is no observable difference in the evolution of the night-side between them and 1.5$R_{\oplus}$ Super-Earths. The day-side however shows a greater quantity of rock formation in the interior of the 1.5$R_{\oplus}$ model. The day-side magma ocean of the 1.5$R_{\oplus}$ model is also much shallower ($\approx 190$ km) than that of the 1.0$R_{\oplus}$ model ($\approx 560$ km). This is because of the higher mass of the 1.5$R_{\oplus}$ planet. With the increase in radius, we also had to increase the mass of the planet. For a 1.5$R_{\oplus}$ planet with a CMF of 32 per cent, the mass was found to be $\approx$ 7.0$M_{\oplus}$, which contributes to a higher gravitational potential than in the 1.0$R_{\oplus}$ planet. Therefore the pressure acting radially along the mantle would be higher. The higher pressure would necessitate a higher temperature to keep the mantle as a melt at a given point. In our model, the temperature profile falls below the solidus curve for a large portion of the mantle because of the aforementioned reasons, leading to more rock formation in the day-side interior. 

As was done with the 1.0$R_{\oplus}$ model, we investigated the variation in $T_{s}$ of the night-side with respect to horizontal heat transport where we changed the value of $Q_{H}$ by varying the magma viscosity. The model showed that the night-side surface solidifies in about 1.2 Gyrs in the absence of $Q_{H}$ and $Q_{T}$. Once we activated $Q_{H}$, the values of night-side $T_{s}$ at the same viscosities used for the 1.0$R_{\oplus}$ model were slightly higher in the 1.5$R_{\oplus}$ model. The reason for this feature can be found in Equation 24. The magnitude of $T_{s}$ on the night-side depends on the gravity and planetary radius, and scales approximately as $g^{\frac{2}{7}} \times R_{p}^{\frac{-15}{28}}$. When we apply the difference in $g$ and $R_{p}$ between a Earth-sized and Super-Earth type planet into this scaling law, we find that the model derived $T_{s}$ for a Super-Earth is approximately 1.05 times the value of $T_{s}$ for a Earth-sized model.

The energy required to keep the night-side melted on a 1.5$R_{\oplus}$ planet is around $5.4 \times 10^{20}$ Watts. Once again this makes sense as the energy radiated from the surface scales with the square of $R_{p}$. The panels in Figure \ref{fig5} demonstrate how the night-side internal structure varies for different magma viscosities. More time is needed for cooling and as a result we see the night-side take approximately 4.8 billion years to fully solidify at $\eta = 1$ Pa s compared to the time taken for the 1.0$R_{\oplus}$ model at the same viscosity. The pattern follows for subsequent values of $\eta$ until we get to $\eta = 10^{-2}$ Pa s. The latter shows that it takes around 10 billion years to fully solidify. For Super-Earth's much younger than this timescale, it is plausible that the night-side would be mush even at $10^{-2}$ Pa s. Between viscosities of $10^{-2}$ and $10^{-4}$ Pa s we can see night-side surfaces of mush, until we get to $\eta = 10^{-5}$ Pa s where we start to see a long-term night-side magma ocean. It is however a lot thinner (about 80 km) compared to the magma oceans possible on the surface of a 1.0$R_{\oplus}$ planet. This is once again due to the larger gravitational potential of a Super-Earth. It is possible that if $Q_{H}$ is the only source of heat powering the night-side, we might see surfaces of mush but no magma oceans on the night-side of Super-Earth's because of the aforementioned reason. This is also assuming that the viscosity cannot be continuously reduced to smaller values. 

Just as we observed in the 1.0$R_{\oplus}$ model, increasing the CMF to 70 per cent did not show a difference from the results with a CMF of 32 per cent. Hence we conclude based on our current findings that the CMF does not affect the thermal evolution, with the parameters we are currently using in our model.

\subsection{The inclusion of tidal heating}   \label{sec3.3}

We added tidal heating $Q_{T}$ to both the 1.0$R_{\oplus}$ and 1.5$R_{\oplus}$ models (\textbf{see Figure \ref{fig4}}). With the constraints we applied on the magnitude of tidal heating from Section 2.3, we initially included only the effect of $Q_{T}$ with $Q_{H}$ set to zero, and then ran a set of simulations with both $Q_{T}$ and $Q_{H}$ given non-zero values. Figure \ref{fig5}a and \ref{fig5}b shows the 2D parameter space of steady-state $Q_{H}$, $Q_{T}$ and the resulting night-side surface conditions along with the results of numerical simulations conducted at different horizontal and tidal heating values. 

For both 1.0$R_{\oplus}$ and 1.5$R_{\oplus}$ planets, at least $8 \times 10^{-4}$ W/kg of tidal heating energy is required to keep the night-side melted if the model relies only on tidal heating without $Q_{H}$. This is approximately $8 \times 10^{5}$ times the amount of tidal heating observed in Io. Otherwise tidal heating alone is not enough to keep the night-side molten. This is because $Q_{T}$ is tied to the mush mass, which increases initially as the magma ocean cools, and then decreases when the mush starts to solidify into rock. As this is not a fixed energy source, the reduction of mush leads to the reduction in tidal heating. The reduction of tidal heating in turn leads to the reduction of the surface temperature, which cycles around to the further reduction of mush. This phenomenon of runaway night-side cooling is why when we only have tidal heating as a heat source for the night-side, we would see high night-side surface temperatures between 100 million and 2 billion years, before it starts to rapidly reduce leading to the solidification of the night-side. 

For 1.0$R_{\oplus}$ planets, even with the inclusion of $Q_{H}$ the amount of $q_{T}$ required for either melting or creating a sea of mush on the night-side remains at or above $7 \times 10^{-4}$ W/kg for magma magma viscosities up to $10^{-3}$ Pa s. At $\eta = 10^{-3}$ Pa s and at lower magma viscosities, the combination of tidal heating and day-night heat transfer leads to a greater parameter space where a night-side that is either molten or mush is possible. Between viscosities of $10^{-4}$ Pa s and $10^{-5}$ Pa s we see a molten night-side even at values of $q_{T}$ as low as $8 \times 10^{-5}$ W/kg. For 1.5$R_{\oplus}$ planets the minimum wattage per unit mass at which melting occurs on the night-side surface is lower than what was observed for 1.0$R_{\oplus}$ planets. Consequently there is a larger parameter space (See Figure \ref{fig5}) in which a molten or mush night-side surface is possible. The reason for this is that a Super-Earth will have a higher mass of mush at a given point in its cooling history than an Earth-sized planet. Since $Q_{T}$ is proportional to the mass of mush, a higher mush mass means more tidal heating for a given amount of $q_{T}$. Because of the higher mass of Super-Earth's, the cooling rate is slower, leading to longer lasting quantities of mush on the night-side. because of these reasons, the night-side of Super-Earth's can have a mushy night-side surface with $3 \times 10^{-4}$ W/kg at a magma viscosity of $10^{-1}$ Pa s, which is not possible for an Earth-sized planet. From Figure \ref{fig5}d it can be seen that a sea of mush is possible for the night-side of a Super-Earth with energies below $1 \times 10^{-4}$ W/kg at $\eta = 10^{-3}$ Pa s and a molten surface at $q_{T} < 5 \times 10^{-5}$ W/kg at $\eta > 10^{-3}$ Pa s. A Super Earth with a magma viscosity of $\eta = 10^{-3}$ Pa s, which is comparable to the viscosity of heated water, could have a magma ocean or an ocean of mush if both tidal heating and horizontal heating is happening together. 

Applying the equations from \cite{Dris2015} to K2-141 b we found what sort of eccentricity would enable the planet to generate the amount of tidal heating needed to have a molten night-side. If the eccentricity is between 0.007 and 0.01, it is possible to have the $Q_{T}$ needed to melt the night-side (see Figure \ref{fig6}).

\section{Discussion and conclusions}    \label{sec4}

Our thermal evolution model shows under what circumstances the night-side of a lava planets can be kept in a molten or mush state. We also found the timescales at which the night-side can cool down from magma to solid rock and the physical parameters that determine these timescales. In addition to considering the effect of planetary radius, we also increased the CMF in some of our models from 32 per cent to 70 per cent to simulate a Super-Mercury. The day-side and night-side evolution showed no change with an increase in CMF for either 1.0$R_{\oplus}$ or 1.5$R_{\oplus}$ planets. The simulations showed that latent heat does not significantly affect the thermal evolution because the growth rates of the mush and solid layers are too slow to boost the magnitude of the latent heat too slow. The latent heat is at most 100 times less than either $Q_{R}$, $Q_{H}$ or $q_{T}$. If the growth rate is fast enough it could potentially have a slight effect on early evolution but not at billion year timescales. 

The solidus and liquidus curves used in this study \citep{Fiq2010, zhang1994} were extrapolated to pressures beyond 150 GPa. We conducted a sensitivity analysis of the extrapolations of these melt curves by generating synthetic data points above 150 GPa, and refitting for the melt curves including the new data. We adjusted the solidus-liquidus equations with the new fitted parameters and re-ran the models. We found that the refitted melt curves do not alter timescales for the thermal evolution of Super-Earths. This study does however motivate the acquisition of new experimental data for silicate melts at extreme pressures.   

For 1.0$R_{\oplus}$ planets, a minimum of $Q_{H} = 2.3 \times 10^{20}$ Watts of heat must be transferred from the day-side to the night-side to keep it molten. This is approximately 20 per cent of the incident stellar flux being transported across hemispheres. The viscosity of magma needed for this quantity of heat to flow in our model is $10^{-4}$ Pa s. Between viscosities of $10^{-1}$ Pa s and $10^{-3}$ Pa s our model showed a night-side that was kept in a mush state for billions of years. At and below $10^{-1}$ Pa s the energy transmitted horizontally is insufficient to to keep the night-side from solidifying. To put these numbers in context, magma at thousands of K has a viscosity of $10^{-3}$ Pa s, similar to that of water \citep{Malosso2022}. In the case of a 1.5$R_{\oplus}$ Super-Earth, the steady-state energy requirement for a night-side magma ocean is $Q_{H} = 5.2 \times 10^{20}$ Watts, or 23 per cent of the stellar instellation flux. The viscosity of magma required to transport this quantity of energy is about the same as for an Earth-sized planet. It should be noted, however, that the magma oceans of the latter are far shallower than that of the former because of the higher gravity of Super-Earth's. Since the magma viscosity had to be lowered to extremely low values (less than $10^{-4}$ Pa s) to sustain a night-side magma ocean in our models, it is likely that horizontal day-night convection alone is insufficient to maintain a molten night-side. 

For a tidally heated lava planet, the energy requirement for night-side melting is the same as for day-night heat transfer (see Figure \ref{fig5}a and 5b). A 1.0$R_{\oplus}$ planet would require $Q_{T} = 1.2 \times 10^{-4}$ W/kg of tidal heating while a 1.5$R_{\oplus}$ planet would need $Q_{T} = 4.8 \times 10^{-5}$ W/kg of energy. But because tidal heating depends on the mush mass, the simulated quantities deviate from the idealized steady-state values. The tidal energy required for a molten night-side is much higher because the mass of mush decreases as the interior cools down. Specific tidal heating of $q_{T} = 8 \times 10^{-4}$ W/kg is necessary for a sustained night-side magma ocean that would last 10 billion years. Values of $q_{T}$ down to $q_{T} = 5 \times 10^{-4}$ W/kg could keep the night-side molten for several billion years before it cools to either mush or solid. Our simulations show that lava planets can experience a brief period of internal warming when mush mass is at a maximum. The night-side magma ocean is sustained for periods between 800 million and 3 billion years in such instances before they experience runaway cooling as the mush becomes more viscous and solidifies. In Super-Earths, it is more likely that the night-side would either stay molten or mushy with a specific tidal heating $q_{T}$. But their magma oceans and mush oceans will be shallower due to the higher gravity of the Super-Earth. It should be noted however that a sustained partially molten interior due to tidal heating is only possible if the eccentricity required to do so is maintained over billions of years. Studying the orbital evolution and the concurrent interior thermal evolution would shed light on this aspect. 

By the same token, the day-side of 1.5$R_{\oplus}$ planets had a thicker solid layer compared to Earth-sized planets which had more mush (see Figures \ref{fig2}c and \ref{fig2}d), again because of the higher gravitational potential of Super-Earth's leads to greater compression of the interior layers. We see this feature on the night-side as well. Magma and mush oceans, if they can be maintained, will be shallower on Super-Earths. 

Applying the equations from \cite{Dris2015}, we find that tidal heating can maintain a molten night-side on a 1.5$R_{\oplus}$ planet for eccentricity of approximately 0.004, within the uncertainty of current orbital constraints \citep{Malavolta2018, Barragan2018, Zie2022}. Precise night-side measurements of K2-141b, and ideally lava planets of different ages, could provide sensitive constraints on the thermal evolution of their deep interior.

\section*{Acknowledgments}

N.B.C. acknowledges support from an NSERC Discovery Grant, a Tier 2 Canada Research Chair, and an Arthur B. McDonald Fellowship. MH would like to thank the Fonds de recherche du Québec for a doctoral fellowship.  The authors also thank the Trottier Space Institute and l'Institut Trottier de recherche sur les exoplanètes for their financial support and dynamic intellectual environment.

\section*{Data availability}

The data underlying this article are available in the article itself and in any online supplementary material that may be made available.



\bibliographystyle{mnras}
\bibliography{example} 

\begin{thebibliography}{}
\makeatletter
\relax
\def\mn@urlcharsother{\let\do\@makeother \do\$\do\&\do\#\do\^\do\_\do\%\do\~}
\def\mn@doi{\begingroup\mn@urlcharsother \@ifnextchar [ {\mn@doi@} {\mn@doi@[]}}
\def\mn@doi@[#1]#2{\def\@tempa{#1}\ifx\@tempa\@empty \href {http://dx.doi.org/#2} {doi:#2}\else \href {http://dx.doi.org/#2} {#1}\fi \endgroup}
\def\mn@eprint#1#2{\mn@eprint@#1:#2::\@nil}
\def\mn@eprint@arXiv#1{\href {http://arxiv.org/abs/#1} {{\tt arXiv:#1}}}
\def\mn@eprint@dblp#1{\href {http://dblp.uni-trier.de/rec/bibtex/#1.xml} {dblp:#1}}
\def\mn@eprint@#1:#2:#3:#4\@nil{\def\@tempa {#1}\def\@tempb {#2}\def\@tempc {#3}\ifx \@tempc \@empty \let \@tempc \@tempb \let \@tempb \@tempa \fi \ifx \@tempb \@empty \def\@tempb {arXiv}\fi \@ifundefined {mn@eprint@\@tempb}{\@tempb:\@tempc}{\expandafter \expandafter \csname mn@eprint@\@tempb\endcsname \expandafter{\@tempc}}}

\bibitem[\protect\citeauthoryear{{Barrag{\'a}n} et~al.,}{{Barrag{\'a}n} et~al.}{2018}]{Barragan2018}
{Barrag{\'a}n} O.,  et~al., 2018, \mn@doi [\aap] {10.1051/0004-6361/201732217}, \href {https://ui.adsabs.harvard.edu/abs/2018A&A...612A..95B} {612, A95}

\bibitem[\protect\citeauthoryear{{Batalha} et~al.,}{{Batalha} et~al.}{2011}]{Batalha2011}
{Batalha} N.~M.,  et~al., 2011, \mn@doi [\apj] {10.1088/0004-637X/729/1/27}, \href {https://ui.adsabs.harvard.edu/abs/2011ApJ...729...27B} {729, 27}

\bibitem[\protect\citeauthoryear{Boukar\'e, Cowan  \& Badro}{Boukar\'e et~al.}{2022}]{Boukare2022}
Boukar\'e C.-Ã.,  Cowan N.,   Badro J.,  2022, \mn@doi [The Astrophysical Journal] {10.3847/1538-4357/ac8792}, 936, 148

\bibitem[\protect\citeauthoryear{{Boukar{\'e}}, {Lemasquerier}, {Cowan}, {Samuel}  \& {Badro}}{{Boukar{\'e}} et~al.}{2023}]{Boukare2023}
{Boukar{\'e}} C.-{\'E}.,  {Lemasquerier} D.,  {Cowan} N.,  {Samuel} H.,   {Badro} J.,  2023, \mn@doi [arXiv e-prints] {10.48550/arXiv.2308.13614}, \href {https://ui.adsabs.harvard.edu/abs/2023arXiv230813614B} {p. arXiv:2308.13614}

\bibitem[\protect\citeauthoryear{{Castan} \& {Menou}}{{Castan} \& {Menou}}{2011}]{Castan2011}
{Castan} T.,  {Menou} K.,  2011, \mn@doi [\apjl] {10.1088/2041-8205/743/2/L36}, \href {https://ui.adsabs.harvard.edu/abs/2011ApJ...743L..36C} {743, L36}

\bibitem[\protect\citeauthoryear{Chao, deGraffenried, Lach, Nelson, Truax  \& Gaidos}{Chao et~al.}{2021}]{chao2021}
Chao K.-H.,  deGraffenried R.,  Lach M.,  Nelson W.,  Truax K.,   Gaidos E.,  2021, Geochemistry, 81, 125735

\bibitem[\protect\citeauthoryear{{Dang} et~al.,}{{Dang} et~al.}{2021}]{Dang2021}
{Dang} L.,  et~al., 2021, {A Hell of a Phase Curve: Mapping the Surface and Atmosphere of a Lava Planet K2-141b}, JWST Proposal. Cycle 1, ID. \#2347

\bibitem[\protect\citeauthoryear{{Driscoll} \& {Barnes}}{{Driscoll} \& {Barnes}}{2015}]{Dris2015}
{Driscoll} P.~E.,  {Barnes} R.,  2015, \mn@doi [Astrobiology] {10.1089/ast.2015.1325}, \href {https://ui.adsabs.harvard.edu/abs/2015AsBio..15..739D} {15, 739}

\bibitem[\protect\citeauthoryear{{Elkins-Tanton}}{{Elkins-Tanton}}{2012}]{Elkins2012}
{Elkins-Tanton} L.~T.,  2012, \mn@doi [Annual Review of Earth and Planetary Sciences] {10.1146/annurev-earth-042711-105503}, \href {https://ui.adsabs.harvard.edu/abs/2012AREPS..40..113E} {40, 113}

\bibitem[\protect\citeauthoryear{{Espinoza} et~al.,}{{Espinoza} et~al.}{2021}]{Espinoza2021}
{Espinoza} N.,  et~al., 2021, {The first near-infrared spectroscopic phase-curve of a super-Earth}, JWST Proposal. Cycle 1, ID. \#2159

\bibitem[\protect\citeauthoryear{Fiquet, Auzende, Siebert, Corgne, Bureau, Tateno  \& Garbarino}{Fiquet et~al.}{2010}]{Fiq2010}
Fiquet G.,  Auzende A.,  Siebert J.,  Corgne A.,  Bureau H.,  Tateno H.,   Garbarino G.,  2010, \mn@doi [Science (New York, N.Y.)] {10.1126/science.1192448}, 329, 1516

\bibitem[\protect\citeauthoryear{{Hughes} \& {Griffiths}}{{Hughes} \& {Griffiths}}{2008}]{Hugh2008}
{Hughes} G.~O.,  {Griffiths} R.~W.,  2008, \mn@doi [Annual Review of Fluid Mechanics] {10.1146/annurev.fluid.40.111406.102148}, \href {https://ui.adsabs.harvard.edu/abs/2008AnRFM..40..185H} {40, 185}

\bibitem[\protect\citeauthoryear{{Kervazo}, {Tobie}, {Choblet}, {Dumoulin}  \& {B{\v{e}}hounkov{\'a}}}{{Kervazo} et~al.}{2021}]{Kerv2021}
{Kervazo} M.,  {Tobie} G.,  {Choblet} G.,  {Dumoulin} C.,   {B{\v{e}}hounkov{\'a}} M.,  2021, \mn@doi [\aap] {10.1051/0004-6361/202039433}, \href {https://ui.adsabs.harvard.edu/abs/2021A&A...650A..72K} {650, A72}

\bibitem[\protect\citeauthoryear{{Kite}, {Fegley}, {Schaefer}  \& {Gaidos}}{{Kite} et~al.}{2016}]{Kite2016}
{Kite} E.~S.,  {Fegley} Bruce J.,  {Schaefer} L.,   {Gaidos} E.,  2016, \mn@doi [\apj] {10.3847/0004-637X/828/2/80}, \href {https://ui.adsabs.harvard.edu/abs/2016ApJ...828...80K} {828, 80}

\bibitem[\protect\citeauthoryear{{Labrosse}}{{Labrosse}}{2003}]{Lab2003}
{Labrosse} S.,  2003, \mn@doi [Physics of the Earth and Planetary Interiors] {10.1016/j.pepi.2003.07.006}, \href {https://ui.adsabs.harvard.edu/abs/2003PEPI..140..127L} {140, 127}

\bibitem[\protect\citeauthoryear{{Labrosse}}{{Labrosse}}{2015}]{Lab2015}
{Labrosse} S.,  2015, \mn@doi [Physics of the Earth and Planetary Interiors] {10.1016/j.pepi.2015.02.002}, \href {https://ui.adsabs.harvard.edu/abs/2015PEPI..247...36L} {247, 36}

\bibitem[\protect\citeauthoryear{Le~Losq \& Baldoni}{Le~Losq \& Baldoni}{2023}]{Losq2023}
Le~Losq C.,  Baldoni B.,  2023, \mn@doi [Journal of Non-Crystalline Solids] {10.1016/j.jnoncrysol.2023.122481}, 617, 122481

\bibitem[\protect\citeauthoryear{{Lebrun}, {Massol}, {Chassefi{\`e}Re}, {Davaille}, {Marcq}, {Sarda}, {Leblanc}  \& {Brandeis}}{{Lebrun} et~al.}{2013}]{Leb2013}
{Lebrun} T.,  {Massol} H.,  {Chassefi{\`e}Re} E.,  {Davaille} A.,  {Marcq} E.,  {Sarda} P.,  {Leblanc} F.,   {Brandeis} G.,  2013, \mn@doi [Journal of Geophysical Research (Planets)] {10.1002/jgre.20068}, \href {https://ui.adsabs.harvard.edu/abs/2013JGRE..118.1155L} {118, 1155}

\bibitem[\protect\citeauthoryear{{L{\'e}ger} et~al.,}{{L{\'e}ger} et~al.}{2009}]{Leger2009}
{L{\'e}ger} A.,  et~al., 2009, \mn@doi [\aap] {10.1051/0004-6361/200911933}, \href {https://ui.adsabs.harvard.edu/abs/2009A&A...506..287L} {506, 287}

\bibitem[\protect\citeauthoryear{Leger et~al.,}{Leger et~al.}{2011}]{Leger2011}
Leger A.,  et~al., 2011, \mn@doi [Icarus] {10.1016/j.icarus.2011.02.004}, 213

\bibitem[\protect\citeauthoryear{{Malavolta} et~al.,}{{Malavolta} et~al.}{2018}]{Malavolta2018}
{Malavolta} L.,  et~al., 2018, \mn@doi [\aj] {10.3847/1538-3881/aaa5b5}, \href {https://ui.adsabs.harvard.edu/abs/2018AJ....155..107M} {155, 107}

\bibitem[\protect\citeauthoryear{Malosso, Zhang, Car, Baroni  \& Tisi}{Malosso et~al.}{2022}]{Malosso2022}
Malosso C.,  Zhang L.,  Car R.,  Baroni S.,   Tisi D.,  2022, \mn@doi [npj Computational Materials] {10.1038/s41524-022-00830-7}, 8, 139

\bibitem[\protect\citeauthoryear{{Nguyen}, {Cowan}, {Banerjee}  \& {Moores}}{{Nguyen} et~al.}{2020}]{Nguyen2020}
{Nguyen} T.~G.,  {Cowan} N.~B.,  {Banerjee} A.,   {Moores} J.~E.,  2020, \mn@doi [\mnras] {10.1093/mnras/staa2487}, \href {https://ui.adsabs.harvard.edu/abs/2020MNRAS.499.4605N} {499, 4605}

\bibitem[\protect\citeauthoryear{{Nguyen}, {Cowan}, {Pierrehumbert}, {Lupu}  \& {Moores}}{{Nguyen} et~al.}{2022}]{Nguyen2022}
{Nguyen} T.~G.,  {Cowan} N.~B.,  {Pierrehumbert} R.~T.,  {Lupu} R.~E.,   {Moores} J.~E.,  2022, \mn@doi [\mnras] {10.1093/mnras/stac1331}, \href {https://ui.adsabs.harvard.edu/abs/2022MNRAS.513.6125N} {513, 6125}

\bibitem[\protect\citeauthoryear{Ohta, Yagi, Hirose  \& Ohishi}{Ohta et~al.}{2017}]{Ohta2017}
Ohta K.,  Yagi T.,  Hirose K.,   Ohishi Y.,  2017, \mn@doi [Earth and Planetary Science Letters] {10.1016/j.epsl.2017.02.030}, 465, 29

\bibitem[\protect\citeauthoryear{{Patel} et~al.,}{{Patel} et~al.}{2023}]{Patel2023}
{Patel} J.~A.,  et~al., 2023, \mn@doi [\aap] {10.1051/0004-6361/202244946}, \href {https://ui.adsabs.harvard.edu/abs/2023A&A...679A..92P} {679, A92}

\bibitem[\protect\citeauthoryear{{Schaefer} \& {Fegley}}{{Schaefer} \& {Fegley}}{2009}]{Schaefer2009}
{Schaefer} L.,  {Fegley} B.,  2009, \mn@doi [\apjl] {10.1088/0004-637X/703/2/L113}, \href {https://ui.adsabs.harvard.edu/abs/2009ApJ...703L.113S} {703, L113}

\bibitem[\protect\citeauthoryear{{Schaefer}, {Wordsworth}, {Berta-Thompson}  \& {Sasselov}}{{Schaefer} et~al.}{2016}]{Sc2016}
{Schaefer} L.,  {Wordsworth} R.~D.,  {Berta-Thompson} Z.,   {Sasselov} D.,  2016, \mn@doi [\apj] {10.3847/0004-637X/829/2/63}, \href {https://ui.adsabs.harvard.edu/abs/2016ApJ...829...63S} {829, 63}

\bibitem[\protect\citeauthoryear{Solomatov}{Solomatov}{2000}]{solo2000}
Solomatov 2000, \mn@doi [Philosophical transactions. Series A, Mathematical, physical, and engineering sciences] {10.1098/rsta.2008.0125}, 366, 4105

\bibitem[\protect\citeauthoryear{{Tachinami}, {Senshu}  \& {Ida}}{{Tachinami} et~al.}{2011}]{Tach2011}
{Tachinami} C.,  {Senshu} H.,   {Ida} S.,  2011, \mn@doi [\apj] {10.1088/0004-637X/726/2/70}, \href {https://ui.adsabs.harvard.edu/abs/2011ApJ...726...70T} {726, 70}

\bibitem[\protect\citeauthoryear{{Tyler}, {Henning}  \& {Hamilton}}{{Tyler} et~al.}{2015}]{Ty2015}
{Tyler} R.~H.,  {Henning} W.~G.,   {Hamilton} C.~W.,  2015, \mn@doi [\apjs] {10.1088/0067-0049/218/2/22}, \href {https://ui.adsabs.harvard.edu/abs/2015ApJS..218...22T} {218, 22}

\bibitem[\protect\citeauthoryear{Xie, Yoneda, Katsura, Andrault, Tange  \& Higo}{Xie et~al.}{2021}]{Xie2021}
Xie L.,  Yoneda A.,  Katsura T.,  Andrault D.,  Tange Y.,   Higo Y.,  2021, \mn@doi [Geophysical Research Letters] {10.1029/2021GL094507}, 48

\bibitem[\protect\citeauthoryear{{Zhang} \& {Herzberg}}{{Zhang} \& {Herzberg}}{1994}]{zhang1994}
{Zhang} J.,  {Herzberg} C.,  1994, \mn@doi [\jgr] {10.1029/94JB01406}, \href {https://ui.adsabs.harvard.edu/abs/1994JGR....9917729Z} {99, 17,729}

\bibitem[\protect\citeauthoryear{{Zieba} et~al.,}{{Zieba} et~al.}{2022}]{Zie2022}
{Zieba} S.,  et~al., 2022, \mn@doi [\aap] {10.1051/0004-6361/202142912}, \href {https://ui.adsabs.harvard.edu/abs/2022A&A...664A..79Z} {664, A79}

\makeatother
\end{thebibliography}




\appendix

\section{Table of data}

\begin{table*}\centering
  \begin{tabular}{c c c c c}
   \hline
    Parameter & Description & Value & Units & Reference  \\ 
     \hline
    \\
  
    $\mathrm{C}$ & specific heat capacity of silicates & 1200 & J $\mathrm{Kg}^{-1} \mathrm{K}^{-1}$ & \cite{Sc2016}   \\
    $\rho$ & Bulk density of silicates & 4000 & Kg $\mathrm{m}^{-3}$ & \cite{Leb2013}  \\
    $\sigma$ & Stefan-Boltzmann constant & $5.67 \times 10^{-8}$ & W $\mathrm{m}^{-2} \mathrm{K}^{-4}$ \\
    $\kappa$ & thermal conductivity of magma & 13 & W $\mathrm{m}^{-1} \mathrm{K}^{-1}$ & \cite{Ohta2017} \\ 
    $\alpha$ & thermal expansivity & $3 \times 10^{-5}$ & $\mathrm{K}^{-1}$ & \cite{Tach2011} \\
    $\kappa_{d}$ & thermal diffusivity & $7.5 \times 10^{-7}$ & $\mathrm{m}^{2} \mathrm{s}^{-1}$ & \cite{Sc2016} \\
    $\eta_{m}$ & viscosity of magma & 0.1 & Pa s & \cite{Boukare2023} \\
    
     \hline
  \end{tabular}
  \label{tab1}
  \caption{Table of constants used in this study.}
   
\end{table*}


\bsp	
\label{lastpage}
\end{document}